\documentclass[12pt,preprint]{aastex} 

\usepackage{soul}
\usepackage{natbib}
\usepackage{pslatex}

\usepackage{bm}

\usepackage{graphicx}

\usepackage{color, colortbl}

\definecolor{Gray1}{gray}{0.8}
\definecolor{Gray2}{gray}{0.4}
\definecolor{Gray3}{gray}{0.2}

\begin{document}

\title{Stochastic Acceleration of Electrons by Fast Magnetosonic Waves in Solar Flares: the Effects of Anisotropy in Velocity and Wavenumber Space}


\author{Peera Pongkitiwanichakul\altaffilmark{1} \& Benjamin D.\ G.\ Chandran\altaffilmark{2}}

\altaffiltext{1}{Department of Astronomy \& Astrophysics, University
  of Chicago, 5640 S. Ellis Ave, Chicago, IL 60637; peeraoddjob.uchicago.edu}

\altaffiltext{2}{Space Science Center and Department of Physics, University of New Hampshire, Durham, NH 03824}

\begin{abstract}
  We develop a model for stochastic acceleration of electrons in solar
  flares. As in several previous models, the electrons are accelerated
  by turbulent fast magnetosonic waves (``fast waves'') via
  transit-time-damping (TTD) interactions.  (In TTD interactions, fast
  waves act like moving magnetic mirrors that push the electrons
  parallel or anti-parallel to the magnetic field).  We also include
  the effects of Coulomb collisions and the waves' parallel electric
  fields.  Unlike previous models, our model is two-dimensional in
  both momentum space and wavenumber space and takes into account the
  anisotropy of the wave power spectrum $F_k$ and electron
  distribution function $f_{\rm e}$. We use weak turbulence theory and
  quasilinear theory to obtain a set of equations that describes the
  coupled evolution of $F_k$ and $f_{\rm e}$. We solve these equations
  numerically and find that the electron distribution function
  develops a power-law-like non-thermal tail within a restricted range
  of energies $E\in (E_{\rm nt}, E_{\rm max})$.  We obtain approximate
  analytic expressions for $E_{\rm nt}$ and $E_{\rm max}$, which
  describe how these minimum and maximum energies depend upon
  parameters such as the electron number density and the rate at which
  fast-wave energy is injected into the acceleration region at large
  scales. We contrast our results with previous studies that assume
  that $F_k$ and $f_{\rm e}$ are isotropic, and we compare one of our
  numerical calculations with the time-dependent hard-x-ray spectrum
  observed during the June 27, 1980 flare. In our numerical
  calculations, the electron energy spectra are softer (steeper) than
  in models with isotropic $F_k$ and $f_{\rm e}$ and closer to the
  values inferred from observations of solar flares.

\end{abstract}

\maketitle

\newpage

\vspace{0.2cm} 
\section{Introduction}
\label{sec:intro}
\vspace{0.2cm} 

Solar flares involve a rapid increase in the number of photons emitted
at energies exceeding~$\sim 10$~keV.  The photon spectra at these
energies are typically
non-thermal~\citep{Lin81,Lin03,Grigis04,Liu09,Krucker10,Krucker11,Caspi10,Ishikawa11},
indicating the presence of non-thermal
electrons~\citep{Brown71,Miller97}.  One of the proposed mechanisms
for generating these energetic electrons is stochastic particle
acceleration~\citep{Eichler79,Miller96,Miller97,Petrosian06,Benz08}.
In stochastic-particle-acceleration (SPA) models, energy is initially
released from the coronal magnetic field by magnetic
reconnection~\citep{Carmichael64,Hirayama74,Kopp76,Tsuneta92,Tsuneta96,Priest2000}. A
portion of the released energy is in the form of plasma
outflows. Downward-directed outflows collide with closed magnetic
loops lower in the corona, generating electromagnetic
fluctuations. These fluctuations interact with electrons
stochastically, accelerating some of the electrons to high energies.

For the purposes of studying fluctuations with lengthscales much
smaller than the flare acceleration region, the acceleration site can
be modeled as a homogeneous, magnetized plasma with a uniform
background magnetic field $\bm{B}_0$.  Electromagnetic fluctuations
with magnetic fluctuations $\delta b\ll B_0$ can then be approximated
as waves in a homogeneous plasma. At wavelengths exceeding the ion
inertial length $v_{\rm A}/\Omega_{\rm p}$, these waves can be
approximated as magnetohydrodynamic (MHD) waves, i.e., Alfv\'{e}n
waves, fast magnetosonic waves (fast waves), slow magnetosonic waves,
and entropy waves. (The quantity $v_{\rm A} = B_0/\sqrt{4\pi \rho}$ is
the Alfv\'en speed, $\rho$ is the mass density, and~$\Omega_{\rm p}$
is the proton cyclotron frequency.)

Of these wave types, fast waves are thought to be the most effective
at accelerating
electrons~\citep{Miller96,Schlickeiser98,chandran03,Selkowitz04,Yan04,Yan08}.
Fast waves are compressive and modify the magnitude of the magnetic
field as they propagate.  These waves act like moving magnetic
mirrors, exerting forces on the electrons, which enables waves and
electrons to exchange energy.  Such interactions are called
transit-time-damping (TTD) interactions, or simply TTD.  In order for
TTD to cause a secular increase in an electron's energy, the electron
and the wave it interacts with must satisfy the resonance condition,
 \begin{equation}
\omega_{kr}-k_\parallel v_\parallel = 0,
\label{eq:Landau_resonance} 
\end{equation}
where $\omega_{kr}$ is the real part of the wave frequency,
$k_\parallel$ is the component of the wavevector $\bm{k}$ parallel to
$\bm{B}_0$, and $v_\parallel$ is the component of the electron's
velocity parallel to $\bm{B}_0$.  The dispersion relation of fast
waves in low-$\beta$ plasmas such as those found in solar flares
(where $\beta$ is the ratio of plasma pressure to magnetic pressure)
is $ \omega_{kr} = k v_{\rm A}$, and so the resonance condition
reduces to
\begin{equation}
v_\parallel = v_{\rm A}/\cos\theta,
\label{eq:resonance2} 
\end{equation} 
where $\theta$ is the angle between $\bm{k}$ and $\bm{B}_0$.  In the
non-relativistic limit, TTD increases only the parallel kinetic energy
$m_{\rm e} v^2_\parallel/2$ of the electrons, where $m_{\rm e}$ is the
electron mass. The same is true of Landau-damping (LD) interactions,
which are mediated by the waves' parallel electric fields.  On the
other hand, Coulomb collisions and possibly others processes (e.g.,
pitch-angle scattering by whistler waves) can convert parallel kinetic
energy into perpendicular kinetic energy, which is an important
process in SPA models, as we discuss further in Section~\ref{sec:fe}.

Although fast waves are initially excited at large wavelengths by the
interaction between reconnection outflows and magnetic loops, the
energy of these fast waves cascades
turbulent wave-wave interactions.  Fast-wave turbulence is similar to
acoustic turbulence, which transfers wave energy from small k to large
k along radial lines in k-space~\citep{Zakharov70,Cho02,Chandran05}.
This turbulent cascade is important, because it is the largest-$k$
fast waves in such turbulent systems that lead to the strongest TTD
interactions~(Miller et al.~1996; see also Equation~(\ref{eq:dzz})
below).  \nocite{Miller96}  Turbulence also introduces disorder or
randomness into the wave field, causing wave-particle interactions to
become stochastic.

In this work, we extend previous SPA models to allow for anisotropy in
both the fast-wave power spectrum and the electron velocity
distribution.  To the best of our knowledge, this is the first time
that both types of anisotropy have been accounted for within a single
SPA model.  In addition to TTD interactions, we account for LD
interactions and Coulomb collisions.  Our treatment of wave-particle
and wave-wave interactions is based on quasilinear theory and weak
turbulence theory.  We describe our model in detail in
Section~\ref{sec:model}.  In Section~\ref{sec:method}, we describe the
numerical method that we use to solve the equations of our model.  We
compare numerical results from our model to results
from~\cite{Miller96} in Section~\ref{sec:miller}. In
Section~\ref{sec:spectrum} we discuss the evolution of the wave power
spectrum in our model.  In Section~\ref{sec:fe} we derive analytic
expressions describing the anisotropy and maximum energy of the
non-thermal tail of the electron distribution function, which we
compare with new numerical results. In Section~\ref{sec:appflare} we
compare one of our numerical calculations with X-ray observations from
the June 27, 1980~flare. We discuss and summarize our principal
findings in Section~\ref{sec:con}.

\vspace{0.2cm} 
\section{Model}
\label{sec:model} 
\vspace{0.2cm}

We model the electron acceleration region as a box located $\sim
20,000$ km above the chromosphere~\citep{Aschwanden07}, filled with a
homogeneous proton-electron plasma pervaded by a uniform magnetic
field 
\begin{equation}
\bm{B}_0 = B_0 \bm{\hat{z}},
\label{eq:B0z}  
\end{equation} 
where $(x, y, z)$ are Cartesian coordinates.
We define the fast-wave power spectrum in the acceleration region~$F(\bm{k})$, abbreviated $F_k$, to be twice the energy per unit mass
per unit volume in $\bm{k}$-space, where $\bm{k}$ is the wavevector. 
The total fast-wave fluctuation energy per unit mass is given by
\begin{equation}
U_{\rm t} = \frac{1}{2} \, \int F_k\, d^3 k.
\label{eq:Ut} 
\end{equation} 
For simplicity, we assume reflectional symmetry,~$F(- \bm{k}) = F(\bm{k})$. 
We take~$F_k$ to evolve in time according to the equation
\begin{equation}
\frac{\partial F_k}{\partial t}= S_k+\left(\frac{\partial F_k}{\partial t}\right)_{\rm turb}+\left(\frac{\partial F_k}{\partial t}\right)_{\rm res}- k^8 \sin^2 (\theta) \nu F_k .
\label{eq:dFdt} 
\end{equation} 
The term
\begin{equation}
S_k  = \left\{ \begin{array}{cc}\displaystyle \frac{4 \dot{E}_0 }{3 \pi^{3/2} k^3_0} \left(\frac{k}{k_0}\right)^2 \exp\left(-\frac{k^2}{k^2_0}\right) \sigma(\theta) & \mbox{ if $0 < t\leq t_{\rm inj}$} \vspace{0.2cm} \\
0 & \mbox{ if $t > t_{\rm inj}$}
\end{array}
\right.
\label{eq:S}
\end{equation}
is a source term representing fast-wave injection from reconnection
outflows, and $k_0$ is the wavenumber at which~$S_k$ peaks. The term $\sigma(\theta)$
determines the $\theta$-dependence of $S_k$, where $\theta$ is the angle
between~$\bm{k}$ and~$\bm{B}_0$. We normalize~$\sigma(\theta)$ so that $0.5 \int_0^\pi
\sigma(\theta) \sin \theta\, d\theta = 1$.  
The quantity $\dot{E}_0$ is then the total wave energy injection
rate per unit mass.
We set
\begin{equation}
\sigma(\theta)  = \left\{ \begin{array}{cl}\displaystyle \frac{3}{2} \sin^2\theta & \mbox{\,\,\,\, 	for model solutions A1, A2, A3, and A4} \vspace{0.2cm} \\
1 & \mbox{\,\,\,\, for model solutions  B, C and D}
\end{array}
\right.
\label{eq:Sigma}
\end{equation}
where the labels A1, A2, A3, A4, B, C, and D refer to numerical calculations
that we will discuss further in Sections~\ref{sec:miller}
through~\ref{sec:appflare}. The parameter values for these solutions
are listed in Table~\ref{table:set}.  We have considered different
values of $\sigma$ because the best value for the modeling turbulence
in flares is not known.  We take the wave injection to last for a
time~$t_{\rm inj}$, where $t_{\rm inj}$ is an adjustable parameter.

\begin{table}[tp]%
\centering%
{\footnotesize
\begin{tabular}{ccccccccc}
&&&&&&&  \\ 
\hline \hline 
\vspace{-0.3cm} 
\\
Model solution & $B_0 $ & $n_{\rm e}$ & $v_{\rm A}$ & $T_{e\,\, \rm initial}$  & $\beta_{\rm e, initial}$ & $\dot{E}_0$ & $\tau_{\rm cas}$ & $t_{\rm inj}$ 
\\ 
& (G) & $(\mbox{cm}^{-3})$ & $(\mbox{cm} \mbox{ s}^{-1})$ & $(\mbox{K})$ & & $(v^2_{\rm A} \Omega_{\rm p})$ & 
$(\Omega_{\rm p}^{-1})$ & $(\Omega_{\rm p}^{-1})$ \\
\vspace{-0.3cm} 
\\
\hline 
\hline 
\vspace{-0.3cm} 
\\
A1 & 500 & $10^{10}$ & $1.1 \times 10^9$ & $3\times 10^6$ &$4.2\times 10^{-4}$ & $2\times10^{-10}$ & $9.8\times10^5 $ & $3\times10^6 $ \\
A2, A3, A4 & 500 & $10^{10}$ & $1.1 \times 10^9$ & $3\times 10^6$ & $4.2\times 10^{-4}$& $1.8\times 10^{-9}$ &  $(3.25-3.3) \times10^5 $ & $3\times10^6$ \\ 
B & 500 & $10^{10}$ & $1.1 \times 10^9$ & $10^6$ & $1.4\times 10^{-4}$& $5\times10^{-10}$ & $7.8\times10^5 $ & $\infty$ \\ 
C & 250 & $3\times 10^9$ & $1.0 \times 10^9$ & $3\times 10^6$ & $5.0\times 10^{-4}$& $1.5\times10^{-10}$ & $1.3\times10^6 $ & $\infty$ \\
D & 150 & $10^9$ & $1.0 \times 10^9$ & $3\times 10^6$ &$4.6\times 10^{-4}$& $1.25\times10^{-11}$ & $4.4\times10^6 $ & $3\times 10^8$\\ 
\vspace{-0.4cm}  
\\
\hline
\end{tabular}
}
\caption{Parameter values in the numerical calculations that we
  analyze in this paper.  $B_0$ and $n_{\rm e}$ are the background
  magnetic field strength and electron density in the solar-flare
  acceleration region, $v_{\rm A}$ is the Alfv\'en speed, $T_{\rm e,
    initial}$ is the initial electron temperature, $\beta_{\rm e,
    initial} = 8\pi n_{\rm e} k_{\rm B} T_{\rm e, initial}/B_0^2$,
  $\dot{E}_0$ is the rate at which fast-wave energy is injected into
  the solar-flare acceleration region per unit (proton) mass per unit
  time, $\tau_{\rm cas}$ is the energy-cascade timescale defined in
  Equation~(\ref{eq:deftaucas}), and $t_{\rm inj}$ is the duration of
  the fast-wave injection.\label{table:set} 
}
\end{table}

The term $(\partial F_k/\partial t)_{\rm turb}$ in Equation~(\ref{eq:dFdt}) is the so-called ``collision integral" in the wave kinetic equation for weakly turbulent fast waves in low-$\beta$ plasmas derived by~\cite{Chandran05,chandran08}, where $\beta = 8\pi p/B_0^2$ and $p$ is the plasma pressure. 
In particular, we set $(\partial F_k/\partial t)_{\rm turb}$ equal to the right-hand side of Equation~(\ref{eq:turb}) of \cite{Chandran05}, with the Alfv\'en-wave power spectrum~$A_k$ in that equation set equal to zero:
\[
\left(\frac{\partial F_k}{\partial t} \right)_{\rm\!\! turb}
 =   
\frac{ 9 \pi \sin^2\theta}{8 v_{\rm A} } \int  \! d^3\!p\,\,d^3\!q\,
\Big[\delta (k-p-q)kq F_p\big(
F_q - F_k\big)
\]
\begin{equation}
\hspace{0.1cm} + \hspace{0.1cm} \delta (k+p-q)k \big(k F_p F_q +
pF_q F_k - qF_p F_k\big)\Big]\, \delta(\bm{k}
- \bm{p} - \bm{q}) .
\label{eq:turb} 
\end{equation} 
We have neglected Alfv\'{e}n waves for simplicity,  but we expect that
their inclusion would not change our conclusions about electron
acceleration by fast waves. This is because superthermal,
super-Alfv\'enic electrons interact with fast waves with~$\theta >
45^\circ$, which interact only weakly with Alfv\'en
waves~\citep{Chandran05,chandran08}.
In weak fast-wave turbulence, waves with
collinear wavevectors $\bm{k}$, $\bm{p}$, and $\bm{q}$ that satisfy
the wavenumber resonance condition $\bm{k} = \bm{p} + \bm{q}$ and
frequency matching condition $k = p + q$ interact to produce a weak
form of wave steepening, which transfers wave energy from small~$k$ to
large~$k$ along radial lines in $\bm{k}$-space. As $\sin \theta$
decreases, fast waves become less compressive, the fast-wave cascade
weakens, and the energy cascade time increases.  This anisotropy is
represented mathematically by the coefficient of~$\sin^2 \theta$ in
Equation~(\ref{eq:turb}).  When $\sigma(\theta) \propto \sin^2\theta$,
the weakening of~$S_k$ at small~$\theta$ combined with the weakening
of the cascade rate at small~$\theta$ causes $F_k$ to become isotropic
\citep{Chandran05}. We have chosen $\sigma(\theta) \propto
\sin^2\theta$ in numerical calculations A1 through A4 in order to compare our
model with a previous SPA model based on an isotropic~$F_k$
\citep{Miller96}. 

The quantity
\begin{equation}
\tau_{\rm cas} = \frac{U_{\rm t}}{\dot{E}_0}
\label{eq:deftaucas} 
\end{equation} 
is the approximate energy cascade timescale at the forcing
wavenumber~$k_0$, near which most of the fast-wave energy is
concentrated. Because the energy cascade timescale is a decreasing
function of~$k$, $\tau_{\rm cas}$ is also approximately the time
required for fast-wave energy to cascade from $k= k_0$ to $k=
\Omega_{\rm p}/v_{\rm A}$.   We list the values of~$\tau_{\rm cas}$
in our numerical calculations in Table~\ref{table:set}. For
these values, we evaluate~$U_{\rm t}$ after the total fast-wave energy
has reached an approximate steady state.

The second-to-last term in Equation~(\ref{eq:dFdt}) is a damping term
representing resonant interactions between electrons and waves with $k
< k_{\rm max}$. We set
\begin{equation}
\left(\frac{\partial F_k}{\partial t}\right)_{\rm res} = 
\left\{\begin{array}{ll}
 2 \gamma^{(e)}_k F_k & \mbox{ if $k < k_{\rm max}$} \\
0 & \mbox{ if $k\geq k_{\rm max}$}
\end{array}
\right.,
\label{eq:damp2}
\end{equation}
where
\begin{equation}
k_{\rm max} = \frac{\Omega_{\rm p}}{3 v_{\rm A}}
\label{eq:defkappa} 
\end{equation} 
is roughly the maximum wavenumber at which the waves can be
approximated as fast waves. At $k \gtrsim k_{\rm max}$, the fast-wave
branch of the dispersion relation transitions to the whistler
branch. We have set $(\partial F_k/\partial t)_{\rm res} = 0$ at $k>
k_{\rm max}$ in order to exclude the contribution of whistler waves to
electron heating and acceleration. Although potentially important, the
role of whistler waves is beyond the scope of this paper. The
quantity $\gamma^{(e)}_k$ is the imaginary part of the wave frequency,
which we determine using quasilinear theory, as described below.

The last term on the right-hand side of Equation~(\ref{eq:dFdt}) is a
hyperviscous dissipation term, which we include in order to model all
dissipation mechanisms operating at $k > k_{\rm max}$. Although we do
not account for the way that electrons are affected by waves at $k>
k_{\rm max}$ in our model, the power that is dissipated by
hyperviscosity corresponds to power that would, in a real plasma, be
available for electron heating and/or acceleration via
whistler-electron interactions.

We take the electron distribution function $f_{\rm e}$ to evolve
according to the equation
\begin{equation}
\frac{\partial f_{\rm e}}{\partial t} = \left(\frac{\partial f_{\rm e}}{\partial t}\right)_{\rm res}+\left(\frac{\partial f_{\rm e}}{\partial t}\right)_{\rm coll}.
\label{eq:electron}
\end{equation}
The first term in Equation~(\ref{eq:electron}) is the rate of change
of $f_{\rm e}$ resulting from resonant interactions with fast waves,
and is the counterpart to the term $(\partial F_k/\partial t)_{\rm
  res}$ in Equation~(\ref{eq:dFdt}).  The last term in
Equation~(\ref{eq:electron}) is the rate of change of $f_{\rm e}$ due
to Coulomb collisions (see Equation~(\ref{eq:coll}) below).  We model
resonant wave-particle interactions using quasilinear theory.  In this
theory, the Vlasov equation is averaged over many wave periods and
wavelengths.  It is assumed that the fluctuations in the electric and
magnetic fields are from small-amplitude waves, and that the imaginary
parts of the wave frequencies are much smaller than the real parts.
The averaged particle distribution function of species $s$, denoted
$f_s$, then evolves according to the equation~\citep{Kennel66,Stix92}
\begin{equation}
\left(\frac{\partial f_s}{\partial t}\right)_{\rm res} =\lim_{L \to
  \infty} \sum_{n=-\infty}^\infty \pi q_s^2
\left(\frac{2\pi}{L}\right)^3 \int \frac{d^3 k}{p_\perp} G p_\perp
\nonumber \delta (\omega_{kr}-k_\parallel v_\parallel-n\Omega_s)|
\psi_{n,k}^{(s)} | ^2 G f_s,
\label{eq:QLT0}
\end{equation}
where
\begin{equation}
\Omega_{s} = \frac{q_s B_0}{m_s \gamma c}
\label{eq:defOmegas} 
\end{equation} 
is the signed, relativistic cyclotron frequency of species~$s$, $q_s$
and $m_s$ are the charge and mass of a particle of species~$s$,
$\gamma= (1 - v^2/c^2)^{-1/2}$ is the Lorentz factor, $c$ is the speed
of light, $p_\parallel$ ($p_\perp$) is the component of the particle
momentum~${\bm p}$ parallel (perpendicular) to~${\bm B}_0$,
$k_\parallel$ ($k_\perp)$ is the component of~${\bm k}$ parallel
(perpendicular) to~${\bm B}_0$, 
\begin{equation}
G=\left(1-\frac{k_\parallel v_\parallel}{\omega_{kr}}\right)\frac{\partial}{\partial p_\perp}
+\left ( \frac{k_\parallel v_\perp}{\omega_{kr}}\right)\frac{\partial}{\partial p_\parallel},
\label{:G}
\end{equation}  
\begin{equation}
\psi_{n,k}^{(s)}=\frac{1}{\sqrt{2}}[E^+_k e^{i \phi}J_{n+1}(z)+E^-_k e^{-i \phi} J_{n-1}(z)]+\frac{p_\parallel}{p_\perp}E_{kz} J_n(z),
\label{:psi}
\end{equation}
$z=k_\perp v_\perp/\Omega_s$, $J_n$ is the Bessel function of
order~$n$, $E^{\pm}_k=(E_{kx} \mp i E_{ky})/\sqrt{2} $, ${\bm E}_k$
(${\bm B}_k$) is the Fourier transform of the electric (magnetic)
field, and $\phi$ is the azimuthal angle in $\bm{k}$-space.  The
quantity~$L$ is the length scale of the window function that multiplies
functions of position before we take a Fourier transform. Our
Fourier-transform convention, described further in Appendix~\ref{appen:reldamping},  differs from that of \cite{Stix92} by
factors of~$2\pi$, which accounts for why the right-hand side of
Equation~(\ref{eq:QLT0}) is a factor of~$(2\pi)^3$ larger than the
right-hand side of Equation (17-41) of \cite{Stix92}.  The species
subscript $s$ is ``p'' for protons or ``e'' for electrons.  The delta function
in Equation~(\ref{eq:QLT0}) implies that strong interactions occur
only when waves and particles satisfy the resonance condition
\begin{equation}
\omega_{kr}-k_{\parallel} v_\parallel=n\Omega_s.
\label{eq:res} 
\end{equation}
TTD and Landau damping arise when the resonance condition with $n=0$
is satisfied. 

To evaluate $|\psi_{n,k}^{(s)}|^2$ and~$\omega_{kr}$, we treat the fast
waves as if they were propagating in a plasma with
the (non-relativistic) bi-Maxwellian distribution function
\begin{equation}
f_{\rm BM} = \frac{n_{\rm e}}{\pi^{3/2} m_{\rm e}^3 v_{\perp \rm
    T}^2 v_{\parallel \rm T}} \exp\left(-\frac{v_\perp^2}{v_{\perp \rm T}^2} -
  \frac{v_\parallel^2}{v_{\parallel \rm T}^2}\right),
\label{eq:fM}
\end{equation} 
where $n_{\rm e}$ is the electron density, 
\begin{equation}
v_{\perp \rm T} = \sqrt{\frac{2k_{\rm B} T_{\perp \rm e}}{m_{\rm e}}}
\qquad v_{\parallel \rm T} = \sqrt{\frac{2k_{\rm B} T_{\parallel \rm e}}{m_{\rm e}}}
\label{eq:defvperpT} 
\end{equation} 
are the perpendicular and parallel electron thermal speeds, and
$T_{\perp \rm e}$ and $T_{\parallel \rm e}$ are the perpendicular and
parallel electron temperatures.  The factor of~$m_{\rm e}^3$ is
included in the denominator of Equation~(\ref{eq:fM}) because we have
defined~$f_{\rm e}$ to be the number of particles per unit volume in
physical space per unit volume in momentum space (i.e., $\int d^3 p
f_{\rm e} = n_{\rm e}$).  After setting~$f_{\rm e} = f_{\rm BM}$, we
expand the hot-plasma dispersion relation in the limit that
$|\omega_{kr}| \ll \Omega_{\rm p}$, $k_\perp v_\perp \ll |\Omega_{\rm
  e}|$ , and $\omega_{kr}/k_\parallel v_{\parallel \rm T} \sim {\cal
  O}(1)$. The details of this procedure are given in Section~4 of
Chapter~11 of \cite{Stix92}.  Fast waves in this limit satisfy
\begin{equation}
\omega_{kr} = k v_{\rm A}.
 \label{eq:disp}
\end{equation}
For the case in which $\bm{k}$ is in the $x-z$ plane, 
\begin{equation}
\frac{i E_{kz}}{E_{ky}} = -\frac{k_\perp k_\parallel v^2_{\perp \rm T}}{2 \omega_{kr} \Omega_{\rm e}},
 \label{eq:polarization}
\end{equation}
$|E_{kx}| \ll |E_{ky}|$, and
\begin{equation}
\psi_{0,k}^{(e)} =  -\, \frac{ik_\perp v_\perp}{2\Omega_{\rm e}} \left(
1 - \frac{v_{\perp \rm T}^2}{v_\perp^2} \right) E_{ky}.
\label{eq:psi0} 
\end{equation} 
We note that Equation~(\ref{eq:polarization}) differs from
Equation~(31) of Chapter~17 of \cite{Stix92}, because the latter
equation only applies when~$v_{\perp \rm T} = v_{\parallel \rm T}$.
Restricting Equation~(\ref{eq:QLT0}) to ``Landau-resonant'' interactions
(i.e., $n=0$),  we rewrite Equation~(\ref{eq:QLT0}) in the form
\begin{equation}
\left(\frac{\partial f_{\rm e}}{\partial t}\right)_{\rm res} = \frac{\partial}{\partial p_\parallel} \left(D_{\rm res} \frac{\partial f_{\rm e}}{\partial p_\parallel}\right),
\label{eq:ttd_diffusion}
\end{equation}
where 
\begin{equation}
D_{\rm res} = \frac{\pi^2 m_{\rm e}^2 \gamma^2 }{4} \frac{(v_\perp^2 - v_{\perp \rm T}^2)^2}{|v_\parallel^3|} \left(1 - \frac{v_{\rm A}^2}{v_\parallel^2}\right)
 \int_{-1}^1 d(\cos\theta)\; \delta\Big(\!\cos\theta - \frac{v_{\rm A}}{v_\parallel}\Big) \int_0^{\rm k_{\rm max}} dk  \,k F_k .
 \label{eq:dzz}
\end{equation}
Ordinarily, the upper limit on the $k$ integration would be~$+
\infty$, as in Equation~(\ref{eq:QLT0}). However, in
Equation~(\ref{eq:damp2}) we have restricted the $k$ integration to~$k
< k_{\rm max} = \Omega_{\rm p}/3v_{\rm A}$, in order to exclude
wave-particle interactions involving whistler waves. We therefore must
do the same in Equation~(\ref{eq:dzz}) in order to maintain energy
conservation.  To express $D_{\rm res}$ in Equation~(\ref{eq:dzz}) in
terms of~$F_k$ instead of~$|E_{ky}|^2$, we have made use of
Equations~(\ref{eq:FW}) and (\ref{eq:WE}) below and 
our assumption of spherical symmetry
about the $z$~axis, which allows us to evaluate the $\phi$~integral in
Equation~(\ref{eq:QLT0}) by taking~$\bm{k}$ to be in the~$x-z$ plane and
then replacing $\int_0^{2\pi} d\phi(\dots)$ with~$2\pi \times (\dots)$.

Equation~(\ref{eq:dzz}) differs from the momentum diffusion
coefficient~$D_p$ given in Equation~(2.2a) of \cite{Miller96} in two
ways. First, $D_{\rm res}$ in Equation~(\ref{eq:dzz}) is the
coefficient for diffusion in~$p_\parallel$, whereas $D_p$ in
Equation~(2.2a) of \cite{Miller96} is the coefficient for diffusion
in~$p$ when rapid pitch-angle scattering isotropizes~$f_{\rm e}$.
Second, Equation~(\ref{eq:dzz}) accounts for LD
interactions mediated by the parallel component of the electric
field,~$E_{kz}$. The parallel electric field is responsible for the
terms proportional to~$v_{\perp \rm T}^2$ in Equations~(\ref{eq:psi0})
and (\ref{eq:dzz}). The minus signs preceding these terms reflect the
fact that the electric force on electrons is $180^\circ$ out of phase
with the $\mu\nabla B$ force on the electrons~\citep{Stix92}. For fast
waves in Maxwellian plasmas, the effects of the parallel electric
field are quite important.  As noted by \cite{Stix92}, the parallel
electric field in a Maxwellian plasma reduces the fast-wave damping
rate by a factor of~2 relative to the case in which $E_{kz}$ is
neglected (i.e., the case in which the fast waves are damped only by TTD). On the
other hand, for electrons with $v_\perp \gg v_{\perp \rm T}$, TTD
interactions are much stronger than LD interactions, and
the parallel electric field leads to only a small reduction in~$D_{\rm
  res}$.

Returning to Equation~(\ref{eq:damp2}), when the imaginary part of the
wave frequency~$\gamma_k$ is much less than the real part, $\gamma_k$
can be determined using quasilinear theory~\citep{kennel67}. In
Appendix~\ref{appen:reldamping}, we show that the general form
of~$\gamma_k$, allowing for relativistic particles and cyclotron
$(n\neq0)$ interactions, is given by $\gamma_{\bm{k}} = \sum_s
\gamma^{(s)}$, where
\begin{equation}
\gamma_{\bm{k}}^{\rm (s)}=  \sum_{n=-\infty}^\infty \frac{\pi^2 q^2_s}{2}  \int^{\infty}_0 d p_\perp\int^{\infty}_{-\infty} d p_\parallel \frac{p^2_\perp c^2}{\sqrt{p^2c^2+m^2_sc^4}} \frac{|\psi_{n,\bm{k}}^{(s)} | ^2}{W_{\bm{k}}}\delta (\omega_{kr}-k_{\parallel} v_\parallel-n\Omega_s) G f_s,
\label{eq:reldamp}
\end{equation}
\begin{equation}
W_{\bm{k}}=\frac{1}{16 \pi}\left [ \textbf{B}_k^* \cdot \textbf{B}_k +\textbf{E}_k^*\cdot\frac{\partial (\omega \underline{\underline{\epsilon}}_{\,h})}{\partial \omega}\cdot \textbf{E}_k \right ] 
\label{eq:defW} 
\end{equation}
is one half the wave energy per unit $k$-space volume divided by
$(2\pi)^3$ (see Equation~(\ref{eq:epsw})), and
$\underline{\underline{\epsilon}}_{\,h}$ is the hermitian part of the
dielectric tensor~$\underline{\underline{\epsilon}}$. Since $F_k$ is twice the fast-wave energy per unit mass per unit volume in~$k$ space (see Equation~(\ref{eq:Ut})),
\begin{equation}
W_{\bm{k}} = \left(\frac{L}{2\pi}\right)^3 \frac{\rho F_k}{4}.
\label{eq:FW} 
\end{equation} 
To evaluate the right-hand side of Equation~(\ref{eq:defW}), we again follow the development in Chapter~11 of~\cite{Stix92} and expand 
$\underline{ \underline{ \epsilon}}$ in the 
 that $|\omega_{kr}| \ll \Omega_{\rm p}$, $k_\perp v_\perp \ll |\Omega_{\rm e}|$
, and $\omega_{kr}/k_\parallel v_{\parallel \rm T} \sim {\cal O}(1)$. For
fast waves in this limit with $\bm{k}$ in the~$x-z$ plane,
\begin{equation}
W_{\bm{k}} = \frac{c^2}{8 \pi v^2_{\rm A}} |E_{ky}|^2. 
\label{eq:WE} 
\end{equation}
Given our assumption of cylindrical symmetry about the~$z$ axis,
we can evaluate~$\gamma_k$ at any~$\bm{k}$ by first rotating~$\bm{k}$
about the $z$ axis until it lies in the~$x-z$ plane, and then making
use of Equations~(\ref{eq:psi0}) and (\ref{eq:WE}).
In the non-relativistic limit, Equation~(\ref{eq:reldamp}) reduces to
the value of~$\gamma_k^{\rm (s)}$ derived by \cite{kennel67}.  If we
set $n=0$ and consider only interactions involving electrons, then
Equation~(\ref{eq:reldamp}) gives the value of $\gamma^{(e)}_k$ in
Equation~(\ref{eq:damp2}). 

As a check on our results, we note that for $n=0$ interactions with
non-relativistic, Maxwellian electrons, Equation~(\ref{eq:reldamp})
yields 
\begin{equation}
\gamma^{(e)}_k= -\,\frac{\pi^{1/2}}{4}\;\frac{k^2_{\perp}v_{\rm A}}{|k_{\parallel}|} \sqrt{\frac{m_{\rm e} \beta_{\rm e}}{m_{\rm p}}} \;\exp\left(- \frac{m_{\rm e}}{\beta_{\rm e} m_{\rm p}\cos^2\theta}\right),
\label{eq:ken67}
\end{equation}
where
\begin{equation}
\beta_{\rm e} = \frac{8\pi n_{\rm e} k_B T_{\rm e}}{B^2_0}.
\label{eq:betae} 
\end{equation}
This expression is equivalent to the fast-wave damping rate for
Maxwellian plasmas derived by
\cite{ginz60} (see also \cite{Petrosian06}). 

To determine the value of the collision term $(\partial f_{\rm e}/\partial t)_{\rm col}$ in Equation~(\ref{eq:electron}), we make the following approximations. First, we neglect electron-proton collisions. We also work in the non-relativistic limit,
setting
\begin{equation}
\bm{v} = \frac{\bm{p}}{m_{\rm e}},
\label{eq:vp} 
\end{equation} 
which is a reasonable simplification because we focus on electron energies~$\lesssim 100 \mbox{ keV}$. The Coulomb collision operator for electron-electron collisions can be written in the form~\citep{rosenbluth57}
\begin{equation}
\left (\frac{\partial f_{\rm e}}{\partial t} \right )_{\rm coll}=- C \nabla_v \cdot \bm{J},
\label{eq:coll}
\end{equation}
where
\begin{equation}
C = \frac{{\bf2}\pi\Lambda  e^4}{m^2_{\rm e}},
\label{eq:defC}
\end{equation}
\begin{equation}
\Lambda = 24 - \ln\left[\left(\frac{n_{\rm e}}{1 \mbox{ cm}^{-3}}\right)^{1/2} \left(
\frac{ k_{\rm B} T_{\rm e}}{1 \mbox{ eV}}\right)^{-1}\right]
\label{eq:Coullog}
\end{equation}
is the Coulomb logarithm,
\begin{equation}
\bm{J} = \frac{2}{m_{\rm e}}f_{\rm e} \nabla_v K_1 -\frac{1}{m_{\rm e}}\nabla_v \nabla_v K_2 \cdot \nabla_v f_{\rm e},
\label{eq:jcoll}
\end{equation}
\begin{equation}
K_1= \int \frac{f_{\rm e}}{U}  d^3 p,
\label{eq:kcoll1}
\end{equation}
\begin{equation}
K_2= \int U f_{\rm e} d^3 p,
\label{eq:kcoll2}
\end{equation}
and $U=|\vec{v}-\vec{v}'|$. To evaluate Equation~(\ref{eq:coll})
numerically would require a number of operations per time step
$\propto N^2_v$, where $N_v$ is the number of velocity grid points in
the numerical calculation.  In order to reduce the number of
operations required, we replace $f_{\rm e}$ in
Equations~(\ref{eq:kcoll1}) and~(\ref{eq:kcoll2}) with a Maxwellian
distribution $f_{\rm M}$ of temperature $T_{\rm e}$.  In numerical
calculations A1 and A2, we keep $T_{\rm e}$ fixed at the initial
electron temperature. (As we will discuss further in Section 4, this
is to compare our model to the model of of \cite{Miller96}.)  In
numerical calculations A3, A4, B, C and D, we pick $T_{\rm e}$ so that
$f_{\rm M}$ and $f_{\rm e}$ have the same total energy. This allows
$T_{\rm e}$ to increase during a flare, as seen in hard X-ray
observations (see, e.g., Figure 3 of Lin et al. 1981.)  In
Appendix~\ref{sec:b}, we estimate the error introduced by our
approximations of $K_1$ and~$K_2$ in numerical calculations A4, B, C,
and D. We find that the maximum error is $\lesssim 6\%$ for $K_1$ and
$\lesssim 18\%$ for~$K_2$.

Using these approximated values of $K_1$ and $K_2$, we can rewrite
Equation~(\ref{eq:coll}) as
\begin{equation}
\left (\frac{\partial f_{\rm e}}{\partial t} \right )_{\rm Coll}=4\pi\Lambda  e^4 n_{\rm e}m_{\rm e} \nabla_p \cdot \left [\frac{\nu_s}{2} \frac{\bm{p}}{p^3} f_{\rm e}+\frac{1}{2}\frac{\nu_\parallel}{p^3} \bm{p}\bm{p}\cdot \nabla_p f_{\rm e} +\frac{1}{4}\frac{\nu_\perp}{p^3} (p^2\mathbb{I} -\bm{p}\bm{p})\cdot\nabla_p f_{\rm e} \right],
\label{eq:coll2}
\end{equation}
where $\mathbb{I}$ is the unit matrix,
\begin{equation}
\nu_s = 2 \chi(x_\beta),
\label{eq:gs}
\end{equation}
\begin{equation}
\nu_\parallel = \frac{\chi(x_\beta)}{x_\beta},
\label{eq:gpar}
\end{equation}
\begin{equation}
\nu_\perp = 2\left[ \left ( 1-\frac{1}{2 x_\beta}\right )\chi(x_\beta)+\chi'(x_\beta) \right ],
\label{eq:gperp}
\end{equation}
\begin{equation}
\chi(x) = \frac{2}{\sqrt{\pi}} \int^{x}_0 t^{1/2} e^{-t} dt ,
\label{eq:psix}
\end{equation}
and
\begin{equation}
x_\beta = \frac{p^2}{2 m_{\rm e} k_B T_{\rm e}}.
\label{eq:xbeta}
\end{equation}

\vspace{0.2cm} 
\section{Numerical Method}
\label{sec:method} 
\vspace{0.2cm} In order to solve for the time evolution of $F_k$ and $f_{\rm
  e}$, we integrate Equations~(\ref{eq:dFdt}) and~(\ref{eq:electron})
numerically.  We use an explicit method to integrate Equation~(\ref{eq:dFdt})
--- the numerical algorithm employed by \cite{Chandran05} with a trivial
extension to account for the damping term $(\partial F_k/\partial t)_{\rm res} =
-2\gamma_{\rm k} F_{\rm k}$. If we were to use an explicit method to integrate
Equation~(\ref{eq:electron}), we would need to make the time step~$\Delta t$
exceedingly small in order to maintain numerical stability. We therefore
integrate Equation~(\ref{eq:electron}) using the implicit biconjugate
gradient-stabilized method~\citep{Vorst03}.  We evaluate $v_{\perp \rm T}$ in
Equation~(\ref{eq:dzz}) by setting $n_{\rm e} v_{\perp \rm T}^2 = \int d^3 p
f_{\rm e} v_\perp^2$.  To simplify the numerical algorithm, we treat the
following quantities as constant within a single time step: the damping
rate~$\gamma_k$ used to calculate $(\partial F_k/\partial t)_{\rm res}$ in
Equation~(\ref{eq:dFdt}), the momentum diffusion coefficient~$D_{\rm res}$ used
to calculate $(\partial f_{\rm e}/\partial t)_{\rm res}$ in
Equation~(\ref{eq:ttd_diffusion}), and the electron temperature~$T_{\rm e}$ in
Equation~(\ref{eq:xbeta}).  After each time step, we update the values of
$T_{\rm e}$ in the collision operator for numerical calculations A3, A4, B, C,
and D, but we keep $T_{\rm e}$ fixed in model solutions A1 and A2, as discussed
further in Section~\ref{sec:miller}. After each time step, we also update the
values of $\gamma_k$ and~$D_{\rm res}$.  To calculate $\gamma_k^{(e)}$
numerically, we use the procedure described in Appendix~\ref{appen:reldamping}
following Equation~(\ref{eq:defI}).  With this approach, our numerical treatment
of wave-particle interactions conserves energy to machine accuracy.

In wavenumber space, we use a logarithmic wavenumber grid in both~$k_\perp$ and
$k_\parallel$ (the components of~$\bm{k}$ perpendicular and parallel
to~$\bm{B}_0$), with $k_{\perp i} = (0.2 k_0 ) 2^{i/4}$ for $i = 0, 1, 2, \dots,
N-1$, $k_{\parallel 0} = 0$, $k_{\parallel j} = (0.2 k_0) 2^{(j-1)/4}$ for $j=
1, 2, 3, \dots, N-1$, and $N = 62$. In all of our calculations, we choose~ the
hyperviscosity coefficient $\nu$ so that dissipation is negligible at
$k\le k_{\rm max} = \Omega_{\rm p}/3v_{\rm A}$ but strong enough at $k> k_{\rm
  max}$ to truncate the cascade.

In momentum space, we use 
a pseudo-logarithmic grid in $p_\perp$ and $p_\parallel$.
In $p_\perp$, cell centers are given by
\begin{equation}
p^g_{\perp i}=\frac{p_0 [e^{\alpha (2i-1)}-1]}{e^{\alpha}-1}
\label{:rhoc}
\end{equation}
and cell boundaries are given by
\begin{equation}
p^l_{\perp i}=\frac{p_0 [e^{\alpha (2i-2)}-1]}{e^{\alpha}-1},
\label{:rhob}
\end{equation}
where $p_0 = 2.02\times10^{-2} m_{\rm e} v_{\rm A}$ and $\alpha = 1.83\times 10^{-2}$ for $i=1,2,...,N_p$.
We choose this grid because it extends to $p^l_\perp = 0$ and has the property that $\Delta p_{\perp i+1} = e^{2\alpha} \Delta p_{\perp i}$, where $\Delta p_{\perp i} = p^l_{\perp i+1}-p^l_{\perp i}$ is the ``bin width" in $p_\perp$. The $p_\parallel$ grid is identical to the $p_\perp$ grid.

Before discretizing Equation~(\ref{eq:electron}), we write this equation in the form
\begin{equation}
\frac{\partial f}{\partial t} = -\nabla \cdot \bm{J}_{\rm tot},
\label{eq:genfe}
\end{equation}
where $\bm{J}_{\rm tot}$ is the total electron flux in momentum
space. We then obtain a set of discrete equations by integrating
Equation~(\ref{eq:genfe}) over each grid cell in momentum space and
applying Gauss's theorem, so that $\partial f/\partial t$ within each
cell is given by the electron fluxes through the faces of the
cell. Except at the edges of the simulated portion of momentum space,
the flux through each cell face appears twice in the calculation: as
an increase in the number of electrons in one cell and an equal and
opposite decrease in the number of electrons in an adjacent
cell. Summing over all cells, we conserve the total particle number,
except for a tiny flow of particles out of the numerical domain at
large momenta.

\vspace{0.2cm} 
\section{Comparison with Miller et al (1996)}
\label{sec:miller} 
\vspace{0.2cm} 

In this section, we compare our model with one of the numerical
solutions from \cite{Miller96}, hereafter ``MLM96.'' In particular, we
compare our results with MLM96's ``Case 4,'' which is based on
Kraichnan's (1965) phenomenology of MHD turbulence.  Since MLM96 only
considered TTD, we set $E_{kz}$ to zero in numerical
calculations A1, A2, and A3 in order to compare with their results.
This has the effect of eliminating the $v_{\perp \rm T}^2$ term in
Equation~(\ref{eq:dzz}). (We retain the parallel electric field and
the $v_{\perp \rm T}^2$ term in Equation~(\ref{eq:dzz}) in model
solutions A4, B, C, and~D) The acceleration region in MLM96's model is
homogeneous and has dimension $L_{\rm f} =10^9 \mbox{ cm}$, volume
$10^{27} \mbox{ cm}^3$, electron density~$n_{\rm e} =10^{10} \mbox{
  cm}^{-3}$, and a uniform background magnetic field of strength
500~G.  The electrons are initially Maxwellian with a temperature of
$3 \times 10^6$~K. As time progresses, the electrons in MLM96's model
undergo Coulomb collisions with a background electron population that
remains at $T_{\rm e} = 3\times 10^6$~K, even though the simulated
electrons are heated and accelerated.  For these parameters,
$\beta_{\rm e}$ (defined in Equation~(\ref{eq:betae})) is~$ 4.16\times
10^{-4}$, the electron thermal speed $v_{T_{\rm e}} = \sqrt{k_B T_{\rm
    e}/m_{\rm e}}$ is initially $0.62\,v_{\rm A}$, and electrons with
energy equal to $20$ keV move at speed 7.8~$v_{\rm A}$.  Fast waves
are not present at the beginning of MLM96's numerical calculations,
but are instead injected at the wave number $k_0 =
1.4\times10^{-3}\Omega_{\rm p} /v_{\rm A}$ from $t=0$ to $ t= t_{\rm
  inj} = 3\times 10^6 \Omega^{-1}_p$ at the rate $\dot{E}_0 = 2
\times10^{-10} v^2_{\rm A} \Omega_{\rm p}$.

As a first comparison between our 2D model and MLM96's isotropic
model, we carry out numerical calculation~A1 in Table~\ref{table:set},
which has the same parameters as MLM96's Case 4 and the same treatment
of collisions (fixed $T_{\rm e}$ in Equation~(\ref{eq:xbeta})). Our
choice of $\sigma(\theta)$ in Equation~(\ref{eq:S}) for this
calculation results in a steady-state inertial-range fast-wave power
spectrum that is independent of $\theta$, as discussed following
Equation~(\ref{eq:turb}).

We find that in numerical calculation~A1 the maximum number of electrons with
energies $>20$ keV, denoted $N_{20, \rm max}$, is $2.5\times10^4$, and the
maximum rate at which electrons are accelerated to energies $>20$ keV, denoted
$R_{20, \rm max}$, is $9.2\times10^4$ s$^{-1}$. These values are, respectively,
$\sim 1600$ and $\sim 1500$ times smaller than the corresponding values in
MLM96's case~4.  Only 20\% of the total energy injected into waves in our
numerical calculation is transferred to electrons, while the remainder is
dissipated by hyperviscosity at large~$k$. As mentioned in Section 2, the energy
dissipated by hyperviscosity in our model serves as a proxy for the amount of
energy that cascades to whistler-scale wavelengths $\gtrsim \Omega_{\rm
  p}/v_{\rm A}$. In a real plasma, this energy would also presumably be
transferred to electrons, but electron heating and acceleration by whistlers is
beyond the scope of our model.  In MLM96's case~4, almost all of the wave energy
is transferred to electrons.  One of the reasons that electron acceleration is
less efficient in our model is that in weak turbulence theory the fast-wave
energy cascade is more rapid than in the simple phenomenological model employed
by MLM96. For example, if $k^2 F_k = c_1 k^{-3/2}$, where $c_1$ is a constant,
Equation~(\ref{eq:dFdt}) leads to a cascade rate that is $\simeq 9$ times larger
than the cascade rate assumed by MLM96 (see Appendix~\ref{appen:wave}) ---
hence, $c_1$ would be smaller in our model in order to achieve the same value
of~$\dot{E}_0$. A second reason that electron acceleration is less efficient in
our model is the anisotropy of~$f_{\rm e}$. Transit-time damping increases only
the parallel kinetic energy $m_{\rm e}v_\parallel^2/2$ of the superthermal
electrons, and thus leads to anisotropic electron distributions in which
$v_\perp^2 < v_\parallel^2$ for most of the electrons. For non-thermal electrons
with $|v_\parallel| \gg v_{\rm A}$, $D_{\rm ttd} \propto \gamma^2
v_\perp^4/|v_\parallel|^3$ (see Equation~(\ref{eq:d_ttd}) below), and thus
transit-time damping is less effective in our model than in models in which
$f_{\rm e}$ is isotropic.

In order to isolate the effects of $f_{\rm e}$ anisotropy on electron
acceleration, we carry out a second numerical calculation (A2 in
Table~\ref{table:set}) in which $\dot{E}_0$ is increased by a factor
of 9 so that the wave amplitudes in our model are roughly the same as
in MLM96's Case 4.  We note that increasing $\dot{E}_0$ reduces the
wave cascade time and causes TTD to start earlier in our
larger-$\dot{E}_0$ calculation than in MLM96's Case~4.  With this
larger value of $\dot{E}_0$, the value of $N_{20, \rm max}$ becomes
$2.1\times 10^6$ and the value of $R_{20, \rm max}$ is $7.2\times
10^6$ s$^{-1}$.  These values are both $\sim 20$ times smaller than
the corresponding values in MLM96's Case~4.  We conclude that $f_{\rm
  e}$ anisotropy reduces the efficiency of electron acceleration by
fast magnetosonic waves by a factor of $\sim 20$ for fixed wave
amplitudes. We note, however, that in model solution A2, only 10\% of
the total energy injected into waves is transferred to electrons.  The
remaining energy cascades to wavenumbers $\gtrsim \Omega_{\rm
  p}/v_{\rm A}$, at which it would, in a real plasma, contribute to
further electron heating and acceleration, but via mechanisms not
included in our model.

\begin{figure}[t]
\centerline{\includegraphics[width=12.cm]{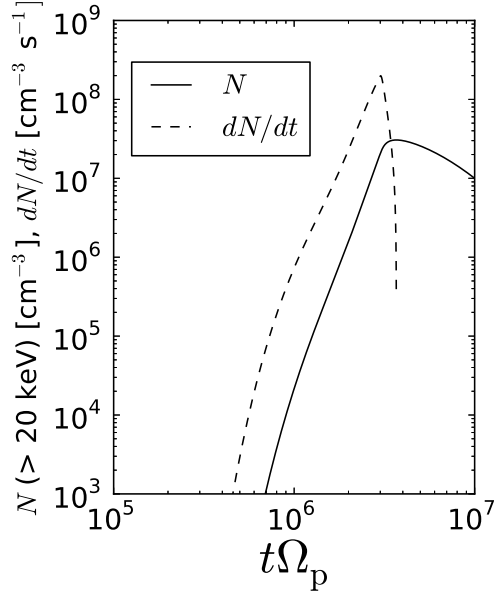}}
\caption{The number~$N$ of electrons  with energies $E$ exceeding 20~keV (solid line) and the acceleration rate~$dN/dt$ in model solution~A3.
At $t= t_{\rm inj}$, wave injection ceases. Subsequently, the waves decay, and $N$ decreases because of Coulomb collisions.
  \label{fig:m20}}\vspace{0.5cm}
\end{figure}

In model solutions A1 and A2, $T_{\rm e}$ is fixed in our approximate
collision operator (Equation~(\ref{eq:xbeta})).  However, as mentioned
previously, $T_{\rm e}$ can increase during a flare.  To investigate
the effect of this increase, we carry out numerical calculation~A3,
which is identical to numerical calculation~A2 except that $T_{\rm e}$
is now allowed to evolve so that $(3/2) n_{\rm e} k_B T_{\rm e}$ is
the total energy density of the instantaneous electron distribution.
In model solution A3, the value of $N_{20 \rm max}$ is $3.1\times 10^7$
and $R_{20 \rm max}$ is $2.0\times 10^8$ s$^{-1}$.  These values are
roughly 15 and 30 time larger than in solution~A2. The reason that
increasing $T_{\rm e}$ in the collision operator enhances the
electron acceleration rate is that the simulated electrons lose less
energy through collisions because they are colliding with
hotter target electrons.  The time evolution of $N_{20 \rm max}$ and
$R_{20 \rm max}$ in solution~A3 are shown in Figure~\ref{fig:m20}.
About 35\% of the total energy injected into waves is transferred to
electrons.

\begin{figure}[t]
\centerline{\includegraphics[width=12.cm]{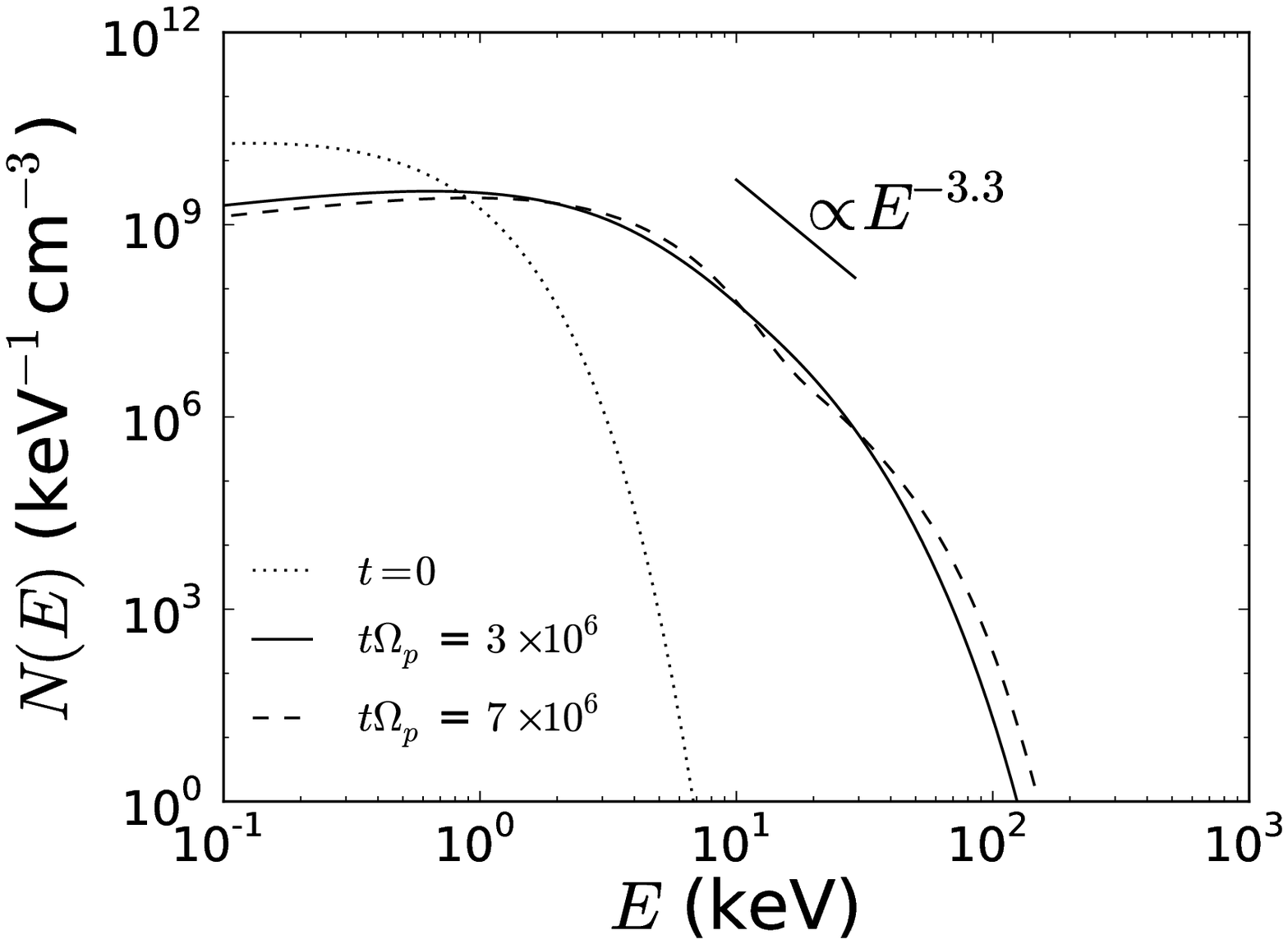}}
\caption{The electron energy spectrum $N(E)$ at three different times in model
solution A3.
  \label{fig:fe}}\vspace{0.5cm}
\end{figure}

In Figure~\ref{fig:fe} we plot the electron energy spectrum
\begin{equation}
N(E) = \frac{2\pi}{c^2} \,p \,\sqrt{p^2 c^2+m^2_{\rm e} c^4}\, \int^1_{-1} d\mu\, f_{\rm e} (p, \mu),
\label{eq:NE}
\end{equation}
in numerical calculation~A3, where $\mu = p_\parallel/p$ and $E
=\sqrt{p^2 c^2+m^2_{\rm e} c^4}-m_{\rm e} c^2$.  As this figure shows,
a power-law-like structure develops over a narrow range of
energies. At the end of the wave-injection period~(i.e., at $t= t_{\rm
  inj} = 3\times 10^6 \mbox{ eV}$), this approximate power law extends
from ~$\sim 7 \mbox{ keV}$ to $\sim 25 \mbox{ keV}$, and $N(E)$ is
roughly proportional to $E^{-3.3}$ in this range, shown in
Figure~\ref{fig:fe}.  A similar power-law-like feature appears in Case
4 of MLM96 (their Figure 11). However, their approximate power law is
much flatter than ours ($\sim E^{-\eta}$ with $\eta$ as small as 1.2)
and extends to larger energies ($> 100 \mbox{ keV})$.

For reference, we carry out a fourth numerical calculation, A4, that
is identical to A3, except that $E_{kz}$ is included. In this
calculation, about 16\% of the total energy injected into waves is
transferred to electrons, which is about half as much as in
solution~A3.  The values of $N_{20 \rm max}$ and $R_{20 \rm max}$ are
$3.7\times 10^{6}$ and $2.0\times 10^7$ s$^{-1}$, respectively. These
values are $\sim$ 8 times and 10 times smaller than those in
solution~A3. These reductions occur for the same reasons that the
inclusion of~$E_{kz}$ reduces the linear damping rate of fast waves in
Maxwellian plasmas by a factor of~2 relative to the case in
which~$E_{kz}$ is neglected~\citep{Stix92}: the parallel electric
force on electrons is~$180^\circ$ degrees out of phase with the
magnetic-mirror force, as discussed in Section~\ref{sec:model}.

\section{Evolution of the Wave Power Spectrum~$F_k$}
\label{sec:spectrum} 

\begin{figure}[t]
\centerline{\includegraphics[width=12.cm]{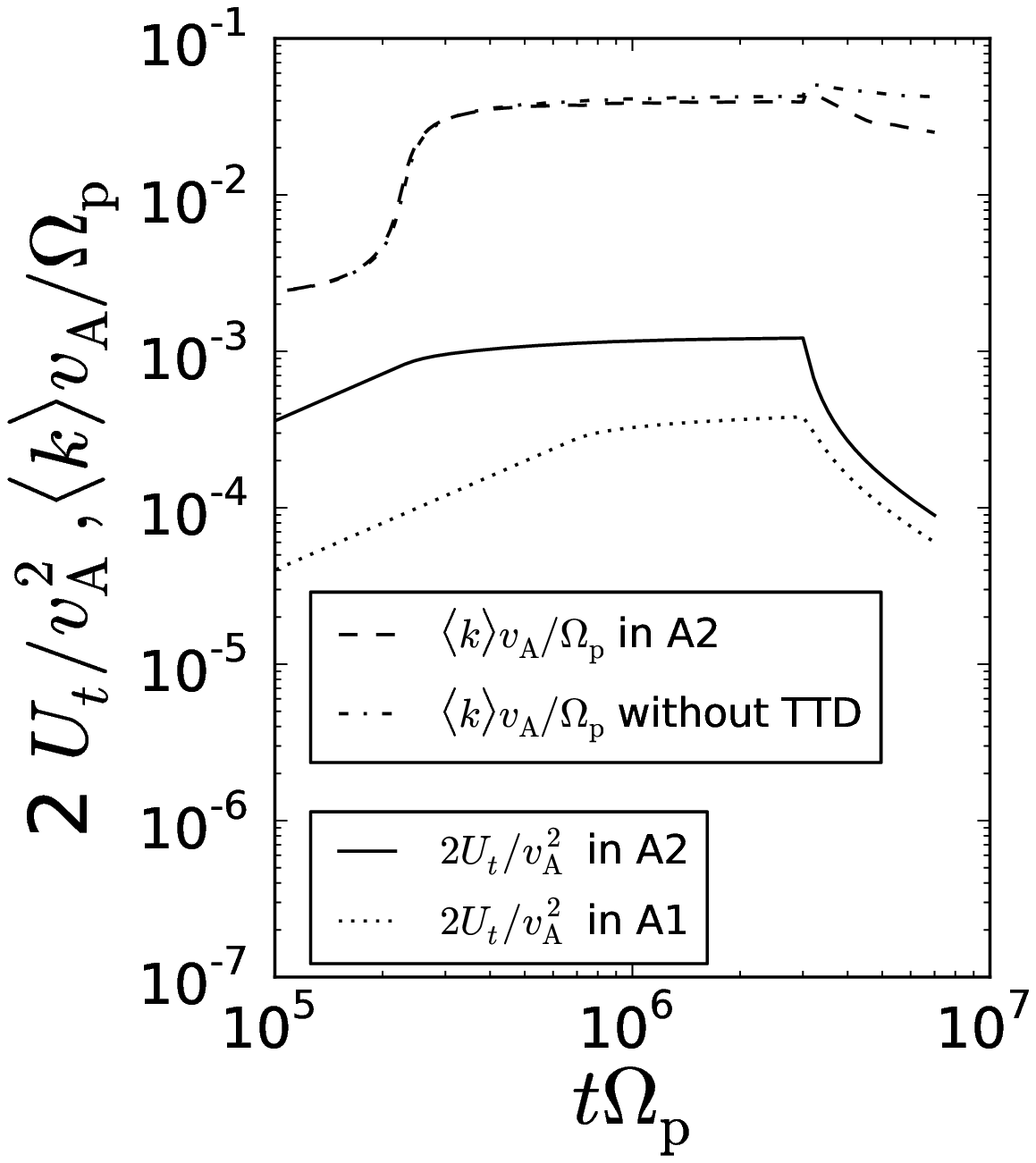}}
\caption{The total wave energy $U_t$ in model solution~A2 (solid line)
  and solution~A1 (dotted line). The dashed line is $\langle k
  \rangle$ in solution~A2, and the dash-dot-dash line is $\langle k
  \rangle$ in a modified version of solution~A2 in which
  transit-time damping is turned off.
 \label{fig:w0}}\vspace{0.5cm}
\end{figure}

In this section, we describe the characteristic way that $F_k$ evolves
in our numerical calculations, using solutions~A1 and~A2 as examples.  In
Figure~\ref{fig:w0}, we plot the energy-weighted average wavenumber
\begin{equation}
\langle k \rangle \equiv \frac{\int d^3k \; k F_k}{\int d^3 k\; F_k}
\label{eq:avk}  
\end{equation} 
for solution~A2 (dashed line) and for a modified version of
solution~A2 in which transit-time damping is turned off (dash-dot-dash
line).  In this modified version of numerical calculation~A2, the value of
$\langle k \rangle$ is somewhat larger than in the original
solution~A2, consistent with the fact that TTD preferentially
removes fast-wave energy at large~$k$.

\begin{figure}[t]
\centerline{\includegraphics[width=12.cm]{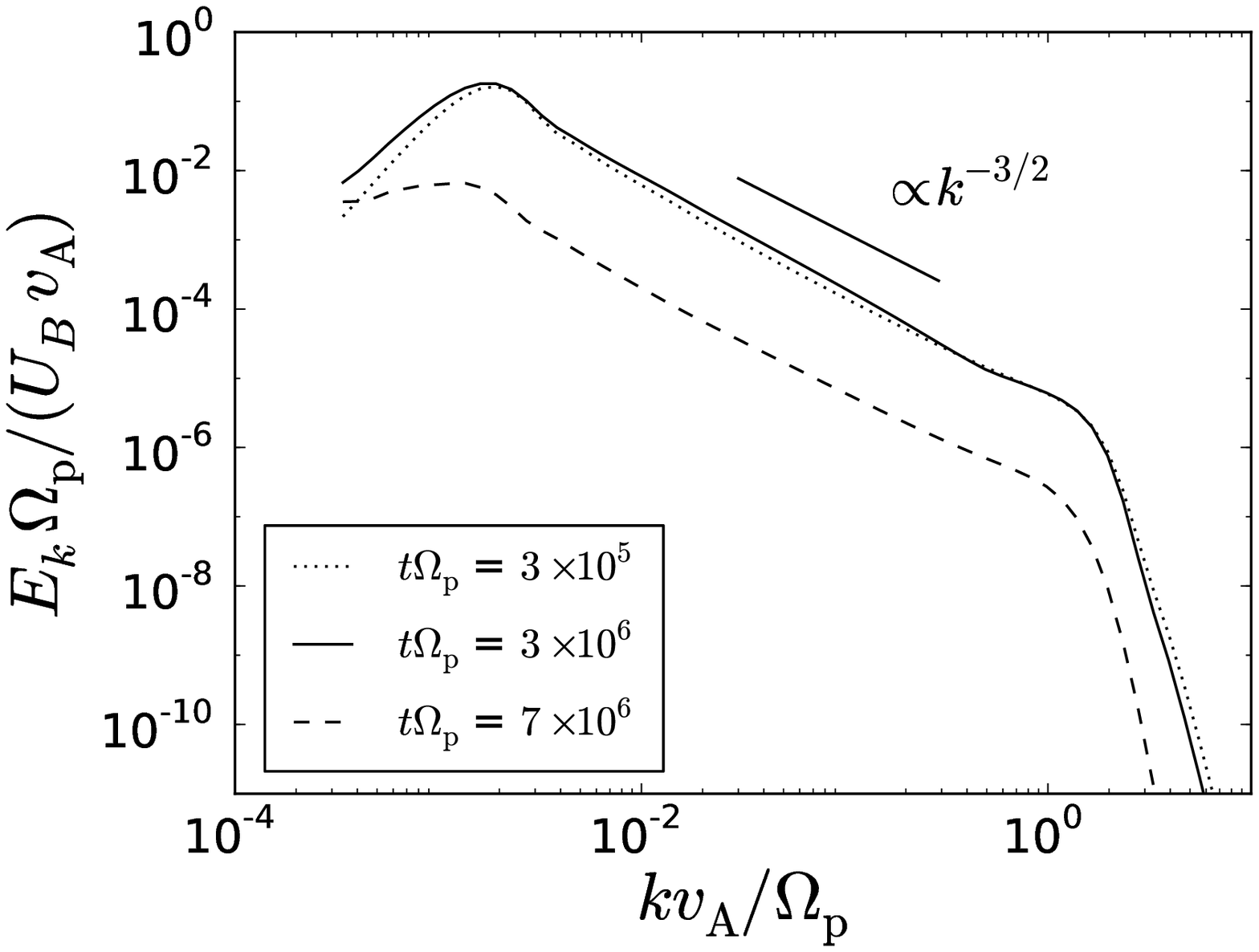}}
\caption{The angle-integrated, $k^2$-compensated fast-wave power spectrum $E_k$ at three different times in model solution A3.
  \label{fig:fk}}\vspace{0.5cm}
\end{figure}

In Figure~\ref{fig:w0} we also plot the total fast-wave fluctuation
energy $U_t$ in numerical calculations~A1 and~A2.  In
Figure~\ref{fig:fk} we plot the angle-integrated, $k^2$-compensated power spectrum
\begin{equation}
E_k = 2\pi \int_0^{\pi} d\theta \, \sin(\theta) k^2 F_k 
\label{eq:defWk} 
\end{equation} 
in numerical calculation~A3 at three different times.  At early times,
$U_t$ grows, but this growth saturates while wave energy is still
being injected. The reason for this saturation is that $F_k$
approaches a state in which energy injection at small~$k$ is balanced
by energy dissipation at large~$k$.  At early times, $\langle k
\rangle$ also grows, as $F_k$ evolves towards a broad power-law-like
spectrum. As can be seen in Figure~\ref{fig:w0}, $U_t$ reaches its
maximum value at an earlier time in solution~A2 than in
solution~A1. This is because the larger values of~$\dot{E}_0$ and
$F_k$ in solution~A2 reduce the energy cascade timescale at the
forcing wavenumber~$k_0$.

As mentioned in Section~\ref{sec:miller}, less than half of the energy
that is injected into waves in all previously described numerical
calculations is transferred to the electrons, and more than half
cascades to $k> k_{\rm max}$ where it is dissipated by hyperviscosity.
We note that much of the wave energy that cascades to $k> k_{\rm max}$
in our numerical calculations is in highly oblique waves with
comparatively large values of $\sin \theta$. There are two reasons for
this.  As discussed in Section~\ref{sec:model}, the energy cascade
time in fast-wave turbulence decreases as $\sin \theta$ increases.  In
addition, because of the TTD resonance condition, waves with $\sin
\theta \sim 1$ interact with only a small number of high-speed
electrons, and thus experience comparatively little damping.

\vspace{0.2cm} 
\section{The Anisotropic Electron Distribution Function}
\label{sec:fe} 
\vspace{0.2cm}

\begin{figure}[t]
\centerline{\includegraphics[width=17.cm]{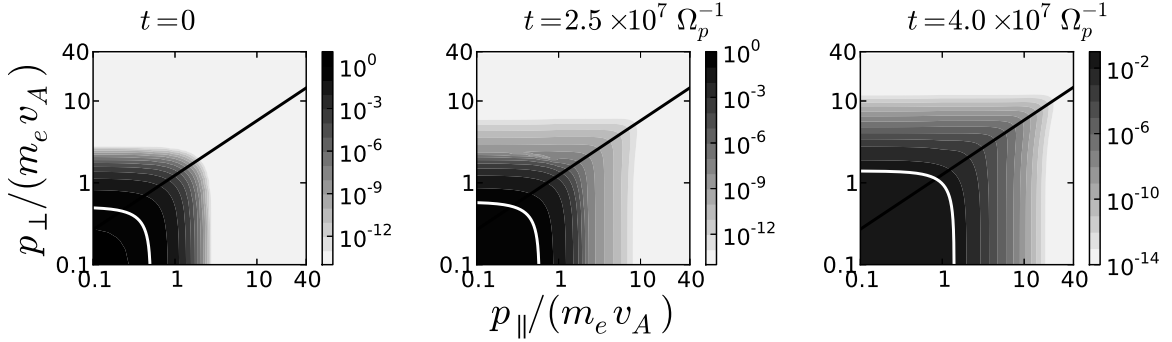}}
\caption{Grey-scale plot of the distribution function $f_{\rm e}$ in
  model solution~B at $t = 0$, $2.5\times10^7 \Omega^{-1}_p$, and
  $4.0\times10^7 \Omega^{-1}_p$.  The solid lines are plots
  Equation~(\ref{eq:blackcurve}), which represents the condition that
  the TTD timescale $\tau_{\rm ttd}$ equals the collisional timescale
  $\tau_{\perp \rm col}$.
\label{fig:fec2}}\vspace{0.5cm}
\end{figure}

In this section, we focus on how resonant wave-particle interactions
and Coulomb collisions affect the anisotropic electron distribution
function. We begin with an example, solution~B of
Table~\ref{table:set}, in which $t_{\rm inj} = \infty$, so that
wave-injection is never shut off.  Figure~\ref{fig:fec2} shows~$f_{\rm
  e}$ at three different times in this numerical calculation.  In the
middle and right panels of this figure, and at a fixed~$p$, $f_{\rm
  e}$ peaks at a pitch angle corresponding approximately to the black
line. (We discuss the precise way in which this black line is
determined later in this section.) The electron distribution becomes
increasingly anisotropic at higher energies, in the sense that the
value of $p_\parallel/p_\perp$ along the black line increases as
$p_\parallel$ increases.  As we will argue in this section, the
anisotropic structure of~$f_{\rm e}$ reflects a balance between
resonant interactions, which accelerates electrons to
larger~$|p_\parallel|$, and collisions, which isotropize the
distribution. For reference, we plot the curve $p=p_{\rm T}$ (white
quarter circles) in Figure~\ref{fig:fec2}, where
\begin{equation}
p_{\rm T} = \sqrt{2 k_B m_{\rm e} T_{\rm e} }
\label{eq:defpt} 
\end{equation} 
is the thermal momentum.

To describe the interplay between wave-particle interactions and
collisions analytically, we begin by obtaining an approximate analytic
expression for the momentum diffusion coefficient~$D_{\rm res}$ in
Equation~(\ref{eq:dzz}). Although some fast-wave energy at $k< k_{\rm
  max} = \Omega_{\rm p}/3v_{\rm A}$ is transferred to electrons via
wave-particle interactions, we make the approximation that most of the
fast-wave energy injected at small wavenumbers cascades to $k> k_{\rm
  max}$, as in the numerical calculations described in
Section~\ref{sec:miller}.  We then model $F_k$ at $k< k_{\rm max}$
using weak-turbulence theory, neglecting losses of fast-wave energy
due to wave-particle interactions.  If fast-wave energy (per unit
mass) is injected into the turbulence isotropically at small~$k$ at
rate~$\dot{E}_0$ (i.e., $\sigma = 1$ in (\ref{eq:Sigma})), then at
$k_0 \ll k < k_{\rm max}$
\begin{equation}
F_k = \left( \frac{4v_{\rm A} \dot{E}_0}{9 \pi^3 c_2}\right) \frac{k^{-7/2}}{\sin\theta},
\label{eq:Fkapprox} 
\end{equation} 
where  $ c_2 = \int_0^\infty
dx\,\ln(1+x)[x(1+x)]^{-5/2}[(1+x)^{9/2} - x^{9/2} - 1] \simeq
26.2$~\citep{Chandran05}. We discuss Equation~(\ref{eq:Fkapprox}) further
in Appendix~\ref{appen:wave}.
We restrict our discussion to superthermal electrons,
setting 
\begin{equation}
p_{\perp} \gg p_{\perp \rm T},
\label{eq:vperplim} 
\end{equation} 
which implies that TTD interactions dominate over LD interactions, as
discussed following Equation~(\ref{eq:dzz}).  Upon substituting
Equations~(\ref{eq:Fkapprox}) and~(\ref{eq:vperplim}) into
Equation~(\ref{eq:dzz}), we find that
\begin{equation}
D_{\rm res} \simeq D_{\rm ttd}
 \label{eq:d_ttd0}
\end{equation}
where
\begin{equation}
D_{\rm ttd} = \left(\frac{\pi}{9c_2}\right)^{1/2} \frac{m_{\rm e}^2 \gamma^2 v_\perp^4}{v_\parallel^4}\left[ k_{\rm max} v_{\rm A} \dot{E}_0 (v_\parallel^2 - v_{\rm A}^2)\right]^{1/2}
\label{eq:d_ttd} 
\end{equation} 
is the parallel-momentum diffusion coefficient arising from TTD
interactions.  We henceforth restrict our analysis to non-relativistic
or trans-relativistic electrons, setting
\begin{equation}
\gamma \simeq 1.
\label{eq:gamma1} 
\end{equation} 
The characteristic timescale on which TTD changes an electron's
parallel momentum by a factor of order unity is
\begin{equation}
\tau_{\rm ttd} \sim \frac{p_\parallel^2}{D_{\rm ttd}}.
\label{eq:deftauttd} 
\end{equation} 

We define the perpendicular (parallel) collisional
timescale $\tau_{\perp \rm col}$ ($\tau_{\parallel \rm col}$) to be
the characteristic time required for Coulomb collisions to change
$p_\perp$ ($p_\parallel$) by a factor of order unity.  At $p\gg p_{\rm
  T}$ and below the black line in Figure~\ref{fig:fec2}, $p_\perp \ll
p_\parallel$. In this region, $p_\perp$ can change by a factor of
order unity when an electron's pitch angle changes by much less than
one radian, which causes $\tau_{\perp \rm col}$ to be $ \ll
\tau_{\parallel \rm col}$. We can show from Equation~(\ref{eq:coll2})
that when $p\gg p_{\rm T}$ and $|p_\parallel| \gg p_\perp$, the
momentum diffusion coefficient for diffusion in~$p_\perp$ is
approximately
\begin{equation}
D_{\perp \rm col} \simeq \frac{\nu_0 (m_{\rm e} v_{\rm A})^3}{ 2|p_\parallel|},
\label{eq:perp_coll}
\end{equation}
where
\begin{equation}
\nu_0 = \frac{4 \pi \Lambda e^4 n_{\rm e}}{m_{\rm e}^2 v_{\rm A}^3}
\label{eq:defnu0} 
\end{equation} 
is the characteristic collision frequency for electrons with
momentum~$m_{\rm e}v_{\rm A}$.  The perpendicular collision timescale
is then
\begin{equation}
\tau_{\perp \rm col} \sim \frac{ p_\perp^2}{D_{\perp \rm col}}.
\label{eq:tauperpcol} 
\end{equation}

The relative importance of TTD and collisions can be determined by comparing
$\tau_{\rm ttd}$ and~$\tau_{\perp \rm col}$. Since $D_{\rm ttd} \propto
v^4_\perp$ for electrons with~$\gamma \simeq 1$, and since Coulomb collisions
become weaker as $v_\perp$ increases, $\tau_{\rm ttd} \ll \tau_{\perp \rm col}$
at sufficiently large~$v_\perp$.  When~$\tau_{\rm ttd} \ll \tau_{\perp \rm
  col}$, electrons diffuse primarily in~$p_\parallel$ rather than~$p_\perp$,
which explains why the contours of constant~$f_{\rm e}$ are horizontal at
large~$p_\perp$ in Figure~\ref{fig:fec2}.  On the other hand, at very
small~$p_\perp$, $\tau_{\perp \rm col} \ll \tau_{\rm ttd}$ and electrons diffuse
in $\ln (p_\perp/p_{\rm T})$ much more rapidly than they diffuse in
$\ln(p_\parallel/p_{\rm T})$. This explains why the contours of constant~$f_{\rm
  e}$ are nearly vertical at small $p_\perp$ in Figure~\ref{fig:fec2}.

The transition between the TTD-dominated regime at large~$p_\perp$ and the collision-dominated regime at small $p_\perp$ occurs when
\begin{equation}
\tau_{\rm ttd} \sim \tau_{\perp \rm col}.
\label{eq:eqtimes} 
\end{equation} 
If we set $\tau_{\rm ttd} = \tau_{\perp \rm col}$, take
$|p_\parallel|$ to be $\gg m_{\rm e} v_{\rm A}$, and replace the $\sim$ signs in
Equations~(\ref{eq:deftauttd}) and (\ref{eq:tauperpcol}) with equals
signs, we obtain
\begin{equation}
\frac{p_\perp}{m_{\rm e}v_{\rm A}} = 1.3 c_3 \left( \frac{v_{\rm A} \nu_0^2}{k_{\rm max}\dot{E}_0}\right)^{1/12} \left(\frac{p_\parallel}{m_{\rm e} v_{\rm A}}\right)^{2/3},
\label{eq:blackcurve} 
\end{equation} 
where $c_3$ is a dimensionless constant, which we have inserted to account for the uncertainties in replacing the $\sim$ signs with $=$ signs. The black  lines
in Figure~\ref{fig:fec2} are plots of Equation~(\ref{eq:blackcurve}) with
\begin{equation}
c_3 = 0.89.
\label{eq:c3} 
\end{equation} 

As mentioned previously, at a fixed $p>p_{\rm T}$, $f_{\rm e}$ reaches
its maximum value close to the black lines in
Figure~\ref{fig:fec2}. To a reasonable approximation, we can thus take
the majority of the electrons at any fixed non-thermal energy~$E$ to
satisfy Equation~(\ref{eq:blackcurve}) to within a factor of order
unity.  In this approximation, we can view all properties of the
non-thermal electrons as functions of the single
variable~$p_\parallel$. For example, $p_\perp = p_\perp(p_\parallel)$,
$\tau_{\rm ttd} = \tau_{\rm ttd}(p_\parallel)$, etc.  The way that
electrons diffuse out to larger energies along the black lines in
Figure~\ref{fig:fec2} is through a combination of two processes. TTD
causes electrons to diffuse in~$p_\parallel$ at a fixed~$p_\perp$, and
Coulomb collisions scatter electrons to larger values of~$p_\perp$.
If we focus on one of the horizontal lines of constant~$f_{\rm e}$
above the black lines in Figure~\ref{fig:fec2}, the
timescale~$\tau_{\rm ttd}$ increases as $p_\parallel$ increases. The
time it takes an electron to reach a point on one of the black lines
in Figure~\ref{fig:fec2} with parallel velocity~$p_\parallel$ is thus
$\sim \tau_{\rm ttd}(p_\parallel)$, or equivalently~$\tau_{\perp \rm
  col}(p_\parallel)$. This timescale is the acceleration timescale,
denoted $\tau_{\rm acc}$:
\begin{equation}
\tau_{\rm acc}(p_\parallel) = \tau_{\perp  \rm col}(p_\parallel).
\label{eq:deftauacc} 
\end{equation} 
With the use of Equations~(\ref{eq:perp_coll}), (\ref{eq:tauperpcol}), and (\ref{eq:blackcurve}), we find that
\begin{equation}
\tau_{\rm acc} = 3.4 c_3^2 \left(\frac{v_{\rm A}}{\nu_0^4 \,k_{\rm max} \dot{E}_0 }\right)^{1/6} \left(\frac{p_\parallel}{m_{\rm e} v_{\rm A}}\right)^{7/3}.
\label{eq:tau_acc} 
\end{equation} 
The largest $|p_\parallel|$ to which an electron can be accelerated,
denoted $p_{\parallel \rm max}$, is
approximately given by the condition
\begin{equation}
\tau_{\rm acc}(p_{\parallel \rm max}) = \Delta t,
\label{eq:tauaccDt} 
\end{equation} 
 where 
\begin{equation}
\Delta t = t - \tau_{\rm cas}
\label{eq:valDelt} 
\end{equation} 
is the duration of the acceleration process.  The values of the energy
cascade timescale $\tau_{\rm cas}$ (defined in
Equation~(\ref{eq:deftaucas})) in our numerical calculations are
listed in Table~\ref{table:set}.  (We note that at $0<t< t_{\rm cas}$,
$F_k$ is still growing, and Equation~(\ref{eq:d_ttd}), which is the
basis of our analysis, does not apply.)  Equation~(\ref{eq:tauaccDt})
leads to a maximum parallel momentum of
\begin{equation}
p_{\parallel \rm max} = 0.59 c_3^{-6/7} \nu_0^{2/7} \left(\frac{k_{\rm max} \dot{E}_0}{v_{\rm A}}\right)^{1/14} \left(\Delta t\right)^{3/7} m_{\rm e}v_{\rm A}.
\label{eq:pparmax} 
\end{equation} 
In the $\gamma \simeq 1$ limit that we have been focusing on, the
maximum energy~$E_{\rm max}$ that electrons can be accelerated to via
TTD is then
\begin{equation}
E_{\rm max} \simeq \frac{[p_\perp(p_{\parallel \rm max})]^2 + p_{\parallel\rm max}^2}{2m_{\rm e}}
\label{eq:Ent} 
\end{equation} 
We note that Equations~(\ref{eq:pparmax}) and (\ref{eq:Ent}) are valid only when
$t_{\rm cas} < t < t_{\rm inj}$.
At larger values of~$t$, after wave injection ceases, the fast-wave energy decays away, TTD interactions cease, and the electrons undergo a purely collisional evolution, which is described further in Section~\ref{sec:appflare}.

\begin{figure}[t]
\centerline{\includegraphics[width=7.cm]{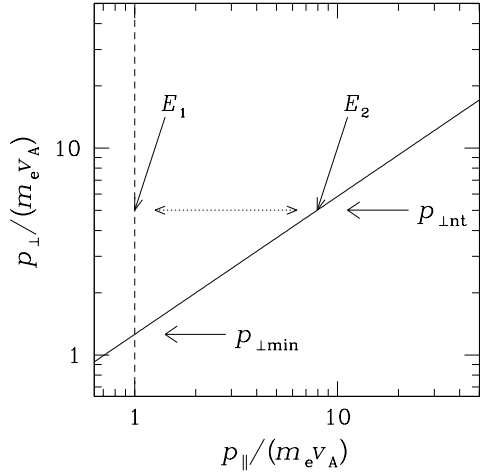}}
\caption{The solid line is the solution to Equation~(\ref{eq:blackcurve}) for
  some arbitrary choice of parameters. This line gives the location in the
  $(p_\parallel, p_\perp)$ plane at which $\tau_{\rm ttd} = \tau_{\perp \rm
    col}$. The vertical dashed line $p_\parallel = m_{\rm e} v_{\rm A}$ shows
  the minimum $p_\parallel$ for which electrons can undergo resonant TTD
  interactions.  The horizontal dotted line represents $p_\parallel$-diffusion
  due to TTD, which 
dominates over collisions above the solid line and to the right of the dashed line. In order for TTD to be dominant, $p_\perp$ must exceed $p_{\perp \rm min}$, which is the $p_\perp$ coordinate of the intersection between the solid and dashed lines. The energies $E_1(p_\perp)$ and $E_2(p_\perp)$ are evaluated, respectively, along the dashed and solid lines.
\label{fig:diagram}}\vspace{0.5cm}
\end{figure}

Referring to Figure~\ref{fig:fe}, the energy~$E_{\rm max}$ is the high-energy cutoff of the non-thermal tail in the electron energy distribution. 
We now discuss,  with the aid of Figure~\ref{fig:diagram}, the physics that determines the minimum energy of this non-thermal tail, which we denote~$E_{\rm nt}$, again restricting our discussion to $t< t_{\rm inj}$.
The vertical dashed line Figure~\ref{fig:diagram}  represents the minimum parallel momentum $p_\parallel = m_{\rm e} v_{\rm A}$ at which electrons can satisfy the TTD resonance condition, Equation~(\ref{eq:resonance2}).

The solid line in this figure is a plot of the solution of
Equation~(\ref{eq:blackcurve}) for some arbitrary choice of
parameters. Above this line, and to the right of the dashed line,
$\tau_{\rm ttd} < \tau_{\perp \rm col}$ and TTD interactions are
dominant. That is, electrons diffuse primarily in $p_\parallel$ rather
than in $p_\perp$, as illustrated schematically with the horizontal
double-headed arrow.  The $p_\perp$ coordinate at the intersection of
the solid and dashed lines is denoted~$p_{\perp \rm min}$ and is the
minimum value of $p_\perp$ for which TTD can dominate over collisions.
Assuming that $E<E_{\rm max}$, electrons with $p_\perp > p_{\perp \rm
  min}$ diffuse rapidly in $p_\parallel$ within the
interval~$p_\parallel \in [m_{\rm e}v_{\rm A}, p_\parallel(p_\perp)]$,
where the function $p_\parallel(p_\perp)$ is obtained by inverting
Equation~(\ref{eq:blackcurve}).  In the non-relativistic limit, the
energies at the endpoints of this $p_\parallel$ interval are
\begin{equation}
E_1(p_\perp) =  \frac{1}{2 m_{\rm e}} \left ( p^2_{\perp}+m^2_{\rm e} v^2_{\rm A} \right )
\label{eq:E1}
\end{equation}
and
\begin{equation}
E_2(p_\perp) =  \frac{1}{2 m_{\rm e}} \left\{ p^2_{\perp}+[p_{\parallel}(p_\perp)]^2 \right\}.
\label{eq:E2}
\end{equation}
These endpoints are labeled $E_1$ and $E_2$ in Figure~\ref{fig:diagram}.
We define the ratio
\begin{equation}
R(p_\perp) = \frac{N_{\rm M}(E_1)}{N_{\rm M}(E_2)} ,
\label{eq:nr}
\end{equation}
where $N_{\rm M}(E)$ is the Maxwellian energy spectrum, obtained by replacing $f_{\rm e}$ in Equation~(\ref{eq:NE})  with $f_{\rm M}$, the Maxwellian distribution that has the same total energy as the instantaneous value of~$f_{\rm e}$.
When $p_\perp$ is just slightly larger than~$p_{\perp \rm min}$, $E_1$ and $E_2$ are not too dissimilar, $R(p_\perp)$ is not very large, and the diffusion of electrons from $p_\parallel = m_{\rm e}v_{\rm A}$ to  $p_\parallel = p_\parallel(p_\perp)$ causes only a minor enhancement of the energy spectrum at $E=E_2$ relative to a Maxwellian energy spectrum. Such a minor enhancement is unable to produce a noticeable non-thermal tail in~$N(E)$. However, as $p_\perp$ increases, $R(p_\perp)$ grows, and 
eventually the diffusion 
of electrons from $p_\parallel = m_{\rm e}v_{\rm A}$ to  $p_\parallel = p_\parallel(p_\perp)$
produces a major enhancement in the value of $N(E)$ at $E=E_2$, leading to the presence of a substantial non-thermal tail in the distribution. In our numerical calculations, we find that the non-thermal tail begins at an energy $\sim E_2(p_{\perp \rm nt})$, where $p_{\perp \rm nt}$ is the solution of the equation 
\begin{equation}
R(p_{\perp \rm nt}) = 100.
\label{eq:R100} 
\end{equation} 
That is,
\begin{equation}
E_{\rm nt} = E_2(p_{\perp \rm nt}).
\label{eq:EntE2} 
\end{equation} 

Qualitatively, there are two main factors that control the value of
$E_{\rm nt}$. The first is the amplitude of the fast-wave
turbulence. As $\dot{E}_0$ and $F_k$ decrease, $p_{\perp \rm min}$
increases, since electrons need larger values of $p_\perp$ for TTD to
dominate over collisions. This causes $E_{\rm nt}$ to increase as a
consequence.  On the other hand, if $\dot{E}_0$ and $F_k$ are
sufficiently large, $E_{\rm nt}$ can be reduced to energies just
moderately above the thermal energy.  The second factor that
influences $E_{\rm nt}$ is the electron temperature. As electrons are
heated, the effects of TTD on~$f_{\rm e}$ become pronounced only at
higher and higher electron energies, causing~$E_{\rm nt}$ to
increase. For example, if at a fixed $p_\perp$, the difference in the
energy between the dashed line and solid line in
Figure~\ref{fig:diagram} is less than $k_{\rm B} T_{\rm e}$, then the
diffusion of electrons from $p_\parallel= m_{\rm e} v_{\rm A}$ to
$p_\parallel = p_\parallel(p_\perp)$ at that value of $p_\perp$ will
have only a minor effect on~$N(E_2(p_\perp))$.  We note that the
location of the black solid lines in Figures~\ref{fig:fec2}
and~\ref{fig:diagram} do not depend upon~$T_{\rm e}$, since~$T_{\rm
  e}$ does not enter into Equation~(\ref{eq:blackcurve}).

\begin{figure}[t]
\centerline{\includegraphics[width=12.cm]{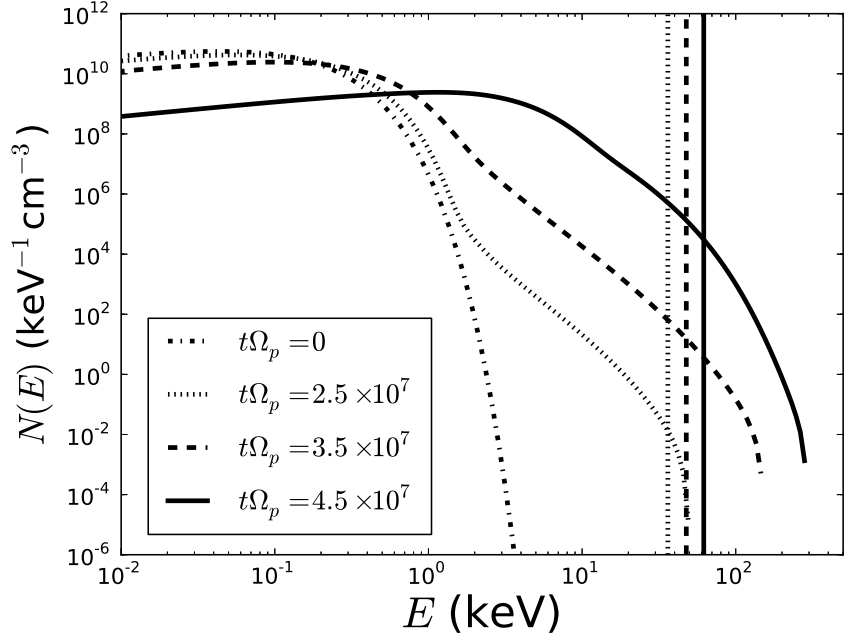}}
\caption{The electron energy spectrum $N(E)$ in model solution~B at
  $t=0$, $2.5\times10^7\Omega^{-1}_p$, $3.5\times10^7\Omega^{-1}_p$,
  and $4.5\times10^7\Omega^{-1}_p$.  The dotted, dashed, and solid
  vertical lines indicate the values of $E_{\rm max}$ from
  Equation~(\ref{eq:Ent}) at $t=2.5\times10^7\Omega^{-1}_p$,
  $t=3.5\times10^7\Omega^{-1}_p$, and $t=4.5\times10^7\Omega^{-1}_p$,
  respectively.
  \label{fig:as}}\vspace{0.5cm}
\end{figure}

\begin{figure}[t]
\centerline{\includegraphics[width=12.cm]{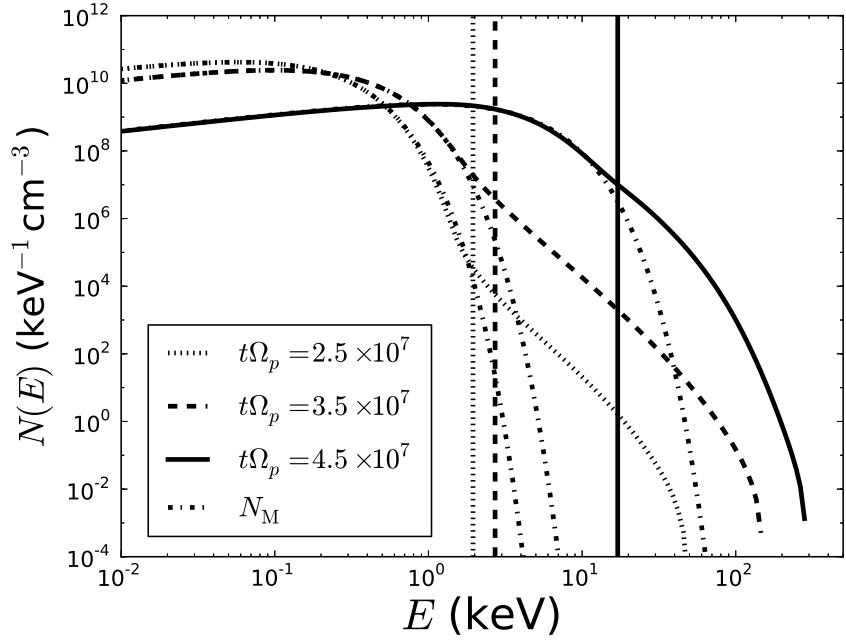}}
\caption{The dotted-line, dashed-line, and solid-line curves are plots
  of $N(E)$ in model solution~B at $t = 2.5\times 10^7 \Omega^{-1}_p$, $t
  = 3.5\times 10^7 \Omega^{-1}_p$, and $t = 4.5\times 10^7
  \Omega^{-1}_p$, respectively. The dash-dot-dash curves are
  Maxwellian energy spectra $N_{\rm M}(E)$ that have the same total
  energy as $N(E)$ at these same three times. The vertical dotted,
  dashed, and solid lines show the values of $E_{\rm nt}$ at the times
  $t = 2.5\times 10^7 \Omega^{-1}_p$, $t = 3.5\times 10^7
  \Omega^{-1}_p$, and $t = 4.5\times 10^7 \Omega^{-1}_p$,
  respectively.
  \label{fig:as2}}\vspace{0.5cm}
\end{figure}

\begin{figure}[t]
\centerline{\includegraphics[width=17.cm]{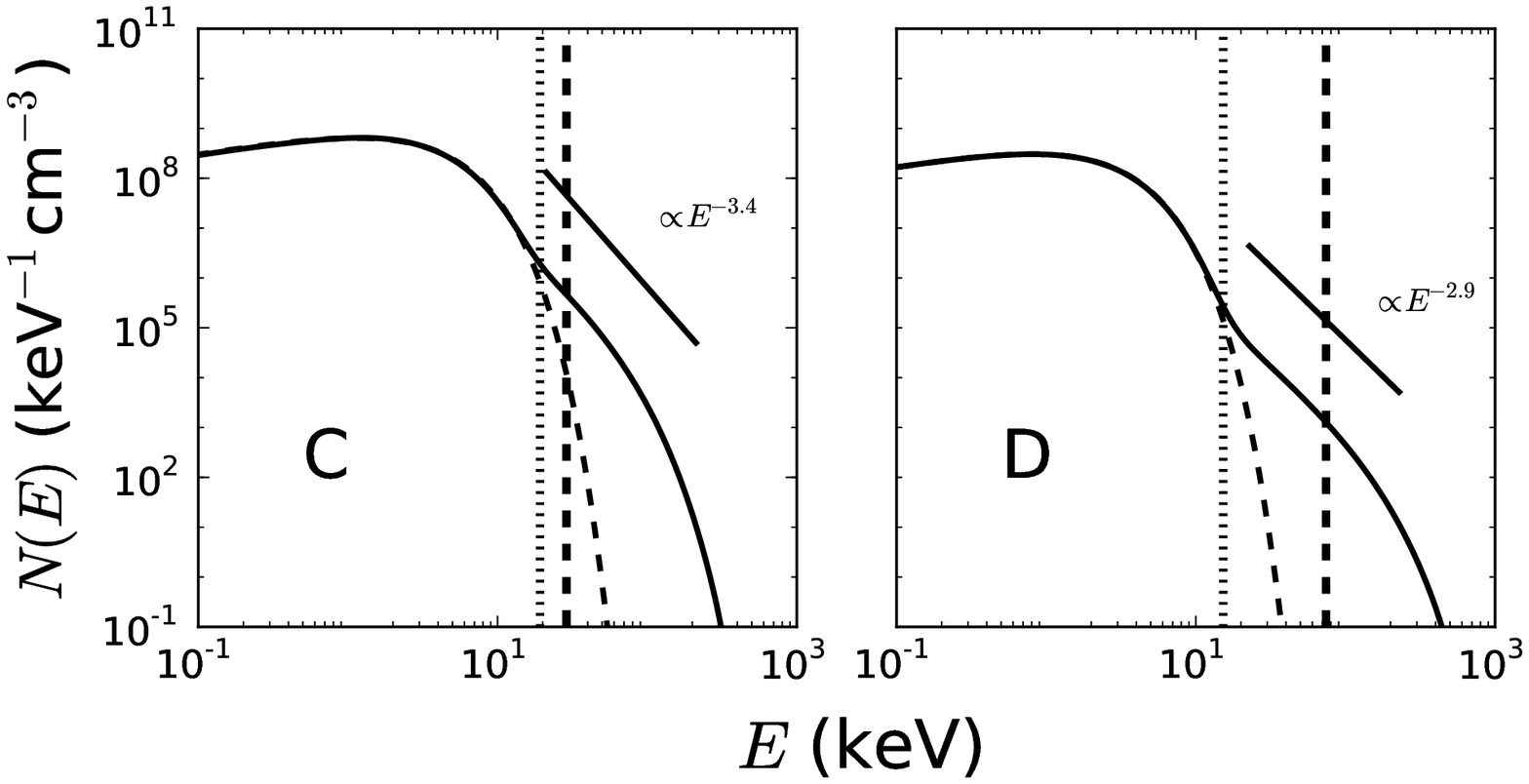}}
\caption{Solid-line curves are $N(E)$ in model solution~C at $t =
  2.8\times 10^7 \Omega^{-1}_p$ (left panel) and solution~D at $t =
  2\times10^8 \Omega^{-1}_p$ (right panel). The thin dashed lines are
  plots of Maxwellian distributions with the same total energy as the
  (non-Maxwellian) electron spectra.  The vertical dotted
  (thick-dashed) lines show the locations of $E_{\rm nt}$ ($E_{\rm
    max}$) in these numerical calculations at these same times.  At
  the moments, the spectral indices are $3.4$ and $2.9$ from the
  solutions C and D, respectively. The straight solid lines are
  power-law fits to $N(E)$ in the energy interval $E_{\rm nt} < E <
  E_{\rm max}$.)
\label{fig:ec2}}\vspace{0.5cm}
\end{figure}

In Figures~\ref{fig:as} and~\ref{fig:as2}, we compare our expressions
for $E_{\rm max}$ and $E_{\rm nt}$ in Equations~(\ref{eq:Ent}) and
(\ref{eq:EntE2}) with the electron energy spectrum in model
solution~B at three different times.
We show the same comparison for two snapshots of solutions C and~D in
Figure~\ref{fig:ec2}.
 For the most part, our expressions for $E_{\rm max}$ and
$E_{\rm nt}$ in Equations~(\ref{eq:Ent}) and (\ref{eq:EntE2})
approximately bound the non-thermal tail in the electron distribution
in our numerical calculations. The least successful fit occurs in
solution~C, for which Equation~(\ref{eq:Ent}) underestimates~$E_{\rm
  max}$ by a factor of~$\sim 2$. A discrepancy of this magnitude,
however, is not entirely surprising, given the approximations we have
made in deriving Equation~(\ref{eq:Ent}).

\vspace{0.2cm} 
\subsection{The Minimum $\beta_{\rm e}$ Required for Efficient~TTD}
\label{eq:betamin} 
\vspace{0.2cm} 

The fraction of the electron population that is significantly affected by TTD depends strongly on $\beta_{\rm e}$. If $\beta_{\rm e} \ll m_{\rm e}/m_{\rm p}$, then the electron thermal speed is much less than~$v_{\rm A}$, and the number of electrons with $|p_\parallel| \gg m_{\rm e} v_{\rm A}$ is exponentially small. 
Since $|p_\parallel|$ must exceed $m_{\rm e} v_{\rm A}$ in order for electrons to satisfy the TTD resonance condition, TTD interactions with fast waves are exceedingly weak if
$\beta _{\rm e} \ll m_{\rm e}/m_{\rm p}$. 

If $\beta_{\rm e}$ is initially small compared to~$m_{\rm e}/m_{\rm p}$ in a flare, 
there may be a transient early stage in a flare in which some process 
heats the electrons until $\beta_{\rm e} \sim m_{\rm e}/m_{\rm p}$. During this initial heating stage, TTD is ineffective at accelerating electrons to non-thermal energies since only a minuscule fraction of the electrons have $p_\parallel > m_{\rm e} v_{\rm A}$. However,
after this heating stage, a significant fraction of electrons satisfy
$p_\parallel > m_{\rm e} v_{\rm A}$, and 
TTD acceleration to higher energies becomes much more efficient.

\vspace{0.2cm} 
\subsection{Power-Law Fits to the Non-thermal Tail}
\label{subsec:index} 
\vspace{0.2cm}

TTD results in a non-thermal tail in the electron energy spectrum that
resembles a power law within the energy range $E_{\rm min} < E <
E_{\rm max}$. We fit the energy spectra in our numerical calculations
within this energy range with a power-law of the form $N(E) \propto
E^{-\eta}$ and show these fits in Figures~\ref{fig:fe}, \ref{fig:ec2},
and~\ref{fig:lin}.  The resulting values of $\eta$ range from $2.9$ to
$3.4$.  As mentioned in Section~\ref{sec:miller}, our electron energy
spectra are steeper than in the isotropic-$f_{\rm e}$ model of
\cite{Miller96}, in which the non-thermal tail in $N(E)$ can scale
like $E^{-\eta}$ with eta as small as 1.2.

\vspace{0.2cm} 
\section{Time Evolution of the Electron Energy Spectrum}
\label{sec:appflare} 
\vspace{0.2cm} 

\begin{figure}[t]
\centerline{\includegraphics[width=13.cm]{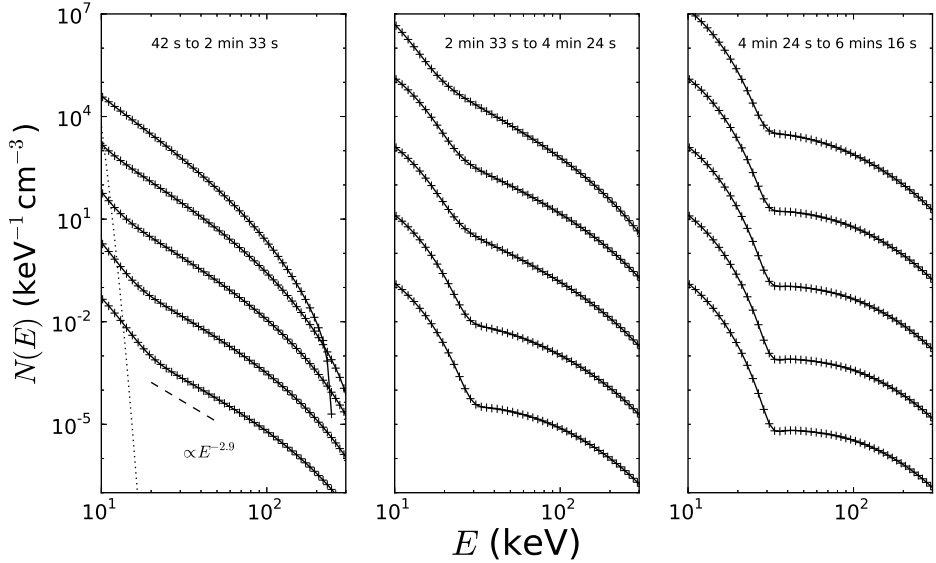}}
\caption{The time evolution of the electron energy spectrum~$N(E)$ in
  model solution~D.  The dotted-line curve in the left panel is the
  initial Maxwellian spectrum at $t=0$~s. The curves plotted with $+$
  signs in all three panels are plots of $N(E)$ at $t=(42 + 27.75 j)
  \mbox{ s}$ with $j=0, 1, 2, ... 12$. These curves are ordered in
  time, from top to bottom in each panel, with $j \in (0,4)$ in the
  left panel, $j\in( 4,8)$ in the middle panel, and $j\in (9,12)$ in
  the right panel. The scale on the vertical axis applies to the
  topmost $+$-sign curve in each panel as well as the dotted line in
  the left panel, with each succeeding plot of~$N(E)$ in each panel
  offset downward by a factor of $10^{-2}$.
\label{fig:lin}}\vspace{0.5cm}
\end{figure}

In Figure~\ref{fig:lin} we plot the electron energy spectrum~$N(E)$ at
different times in model solution~D.  Between $t=0$ and $t\sim 42
\mbox{ s}$, the electron distribution develops a non-thermal,
power-law-like tail extending to $\sim 80 \mbox{ keV}$. As time
progresses, this power-law tail shifts to larger energies, and the
temperature of the thermal particles increases, so that the thermal
distribution shifts into the energy window shown in the figure. After
wave injection ceases at $t=t_{\rm inj} =$ 3~min~23~s, the heating of
the thermal distribution ends, and the non-thermal particles are
gradually pulled back into the thermal distribution by Coulomb
collisions. However, the collision frequency is $\propto p^{-3}$ at
these non-thermal energies, and thus the low-energy end of the
non-thermal tail is affected by collisions earlier than the
high-energy end is affected. As a result, during the collisional
evolution at $t> t_{\rm inj}$, the non-thermal tail drops to lower
amplitudes but becomes flatter, as can be seen in the middle and right
panels of Figure~\ref{fig:lin}.

The evolution of~$N(E)$ shown in Figure~\ref{fig:lin} is qualitatively
similar to the evolution of the hard x-ray spectrum observed in the
June~27, 1980 flare, which is plotted in Figure~3 of \cite{Lin81}. In
both our model and the observations: (1) the power-law part of the
spectrum is confined to a fairly narrow energy range, with $N(E)$
steepening at $E\sim 100 \mbox{ keV}$; (2) the thermal distribution
and non-thermal tail shift to higher energies as time progresses
during the early stages of the flare; and (3) during the late stages
of the flare, the non-thermal tail becomes flatter, but drops in
amplitude.

Although the electron spectrum in solution~D qualitatively resembles
the photon spectrum in the June~27, 1980 flare, our model is not yet
sufficiently sophisticated to produce a synthetic hard x-ray
spectrum~$I(E)$ for a detailed comparison to the observations. In
order for us to map $N(E)$ in our model, which is the electron energy
spectrum in the coronal acceleration region, into an x-ray
spectrum~$I(E)$, we would need to calculate the flux of electrons per
unit energy~$F(E)$ into the chromosphere, and we would need to account
for the way that the escape of particles from the corona
modifies~$N(E)$.

\vspace{0.2cm} 
\section{Discussion and Conclusion}
\label{sec:con} 
\vspace{0.2cm} 

In this paper, we develop a stochastic-particle-acceleration (SPA)
model in which electrons are energized by weakly turbulent fast
magnetosonic waves via a combination of transit-time-damping (TTD)
interactions, Landau-damping (LD) interactions, and pitch-angle
scattering from Coulomb collisions. We use quasilinear theory and weak
turbulence theory to describe the time evolution of the electron
distribution function $f_{\rm e}$ and the fast-wave power spectrum
$F_k$. We solve the equations of this model numerically and find that
TTD leads to power-law-like non-thermal tails in the electron energy
spectrum~$N(E)$ extending from a minimum energy~$E_{\rm nt}$ to a
maximum energy~$E_{\rm max}$.  We derive approximate analytic
expressions for $E_{\rm nt}$ and $E_{\rm max}$ and find that these
expressions agree with our numerical solutions reasonably well.  For a
fast wave, the parallel electric field exerts a force on electrons
that is $180^\circ$ out of phase with the magnetic-mirror force, and
thus the inclusion of the parallel electric field (LD interactions) in
our model reduces the rate of electron acceleration (see, e.g., the
discussion of numerical calculation~A4 at the end of
Section~\ref{sec:miller}).

The main new feature of our model that distinguishes it from previous
studies is our inclusion of anisotropy in both momentum space and
wavenumber space. We assume cylindrical symmetry about the magnetic
field direction in both velocity space and wavenumber space, but allow
$f_{\rm e}$ to depend upon both $p_\perp$ and $p_\parallel$ and $F_k$
to depend on both $k_\perp$ and $k_\parallel$.  Another new feature of
our work in the context of SPA models is our use of weak turbulence
theory to describe the fast-wave energy cascade, which enables us to
avoid introducing an adjustable free parameter into the energy cascade
rate and to account for the weakening of the energy cascade as $\sin
\theta$ decreases, where $\theta$ is the angle between the wavevector
$\bm{k}$ and the background magnetic field~$\bm{B}_0$.

To investigate how much these new features affect our results, we compare one of
our numerical solutions with a numerical example (``Case 4'') published by
\cite{Miller96} (MLM96), which is based on their isotropic SPA model. We find
that there are two main differences between our model and theirs. The first
concerns the energy cascade rate. They modeled the fast-wave energy cascade by
solving a nonlinear diffusion equation for~$F_k$, in which the diffusion
coefficient contained an adjustable free parameter.  If we set the injection
rate to produce $k^2 F_k = A k^{-3/2}$ in both models, where $A$ is some
constant, then the energy cascade rate in our model is roughly 9 times faster
than in their model. Conversely, if we set the energy cascade rates to be equal
in the two models, then $F_k$ is smaller in our model than in theirs by a factor
of~$\simeq 3$, which weakens electron acceleration by fast waves in our model
relative to theirs. The second main difference between the two models is that
the anisotropy of $f_{\rm e}$ reduces the efficiency of electron acceleration
via TTD. This is because TTD accelerates electrons to larger values of
$|p_\parallel|$, but not to larger values of $p_\perp$, which causes most of the
electrons in our model to satisfy $|p_\parallel|> p_\perp$. The TTD momentum
diffusion coefficient $D_{\rm ttd}$ for energetic electrons (with $|v_\parallel|
\gg v_{\rm A}$), however, is $\propto \gamma^2 v_\perp^4/|v_\parallel|^3$, and
the electrons in our anisotropic model thus have smaller values $D_{\rm ttd}$
than in MLM96's isotropic model.  Because of these differences, the total number
of electrons accelerated to energies~$> 20 \mbox{ keV}$ is smaller
in our model than in MLM96's, the power-law-like non-thermal tails in the
electron energy spectrum are steeper in our model, and these tails are limited
to lower maximum energies in our model.

Beyond the comparison with MLM96, our principal results are the following:
\begin{enumerate}
\item In the presence of TTD and Coulomb collisions, the electron
  distribution function at non-thermal energies approaches a specific
  characteristic form, which is shown in Figure~\ref{fig:fec2}. At a
  fixed~$p$, $f_{\rm e}$ peaks at a pitch angle that corresponds to
  the black line in Figure~\ref{fig:fec2}. This line is a plot of
  Equation~(\ref{eq:blackcurve}) and corresponds to the locations in
  the $p_\perp$-$p_\parallel$ plane at which the TTD
  timescale~$\tau_{\rm ttd}$ equals the collisional
  timescale~$\tau_{\perp \rm col}$. Above this black line (at large
  $p_\perp$), TTD dominates over collisions, and rapid
  $p_\parallel$-diffusion of electrons causes $f_{\rm e}$ to become
  almost independent of $p_\parallel$.  Below this curve, collisions
  dominate over TTD, and $f_{\rm e}$ depends more strongly on $\ln
  p_\parallel$ than on $\ln p_\perp$.
\item As can be seen in our expression for $E_{\rm max}$ in
  Equation~(\ref{eq:Ent}), the maximum energy of the non-thermal tail increases
  with increasing electron density~$n_{\rm e}$. This is because collisions help
  electrons to reach higher energies by converting some of the parallel kinetic
  energy ($m_{\rm e} v_\parallel^2/2$) gained via TTD interactions into
  perpendicular kinetic energy ($m_{\rm e} v_\perp^2/2$), which increases the
  rate of TTD acceleration (since $D_{\rm ttd} \propto \gamma^2
  v_\perp^4/|v_\parallel|^3$ for energetic electrons).  Another way of
  thinking about this is that an electron can only reach a point $(p_\parallel ,
  p_\perp)$ on the black line in Figure~\ref{fig:fec2} after the elapsed
  time~$\Delta t$ has grown to a value of order the collisional
  timescale~$\tau_{\perp \rm col}$ at that point (which equals the TTD timescale
  $\tau_{\rm ttd}$ at that point). Consistent with this reasoning, $E_{\rm max}$
  increases with~$t$ up until the wave injection ceases and the waves decay
  away.
\item One of the ways that the magnetic field strength~$B_0$ and the
  initial electron temperature affect electron acceleration via TTD is
  through their influence on the value of~$\beta_{\rm e}$.  If the
  initial value of $\beta_{\rm e}$ is $\ll m_{\rm e}/m_{\rm p}$, then
  only an exponentially small fraction of the electrons have large
  enough values of $|v_\parallel|$ that they can satisfy the TTD
  resonance condition Equation~(\ref{eq:resonance2}), and TTD
  acceleration is exceedingly weak.
\item The time evolution of $N(E)$ in our model solution~D
  qualitatively resembles the time evolution of the hard x-ray
  spectrum~$I(E)$ in the June~27, 1980 flare reported by
  \cite{Lin81}. However, our model is not yet sophisticated enough to
  produce synthetic x-ray spectra, because we have neglected the
  escape of electrons from the acceleration region, which
  alters~$N(E)$ and is needed to determine~$I(E)$.
\end{enumerate}

There are several processes that we have not included in our model.  As just
mentioned, we have not accounted for the escape of electrons from the flare
acceleration region or the flow of low-energy electrons into the acceleration
region from the chromosphere. We have also neglected the escape of fast waves
from the acceleration region (see \cite{Peera12a} for a detailed discussion of
wave escape) and resonance broadening in wave-particle interactions
\citep{shalchi04a,shalchi04b,yan08b,lynn12,lynn13,lynn14}. In the numerical
calculations we have carried out so far, a significant amount of the power
injected into fast waves at small~$k$ cascades to $k>\Omega_{\rm p}/v_{\rm A}$,
where it presumably initiates a cascade of whistler waves. However, our model
neglects the effects of whistler waves and other waves at $k\gtrsim\Omega_{\rm
  p}/v_{\rm A}$ on the electrons. A related point is that we have neglected
non-collisional forms of pitch-angle scattering. One of the effects that waves
at $k>\Omega_{\rm p}/v_{\rm A}$ could have is to enhance the electron
pitch-angle scattering rate. By converting perpendicular electron kinetic energy
into parallel kinetic energy, such enhanced pitch-angle scattering would
increase the efficiency of TTD electron acceleration in flares.

A useful direction for future research would be to incorporate some or
all of these processes into the type of anisotropic SPA model that we
have developed. Another valuable direction for future research would
be to determine the amplitude of fast-wave turbulence in solar flares
using large-scale direct numerical simulations.  Because the
turbulence amplitude plays a critical role in SPA models, a
determination of this amplitude would lead to much more rigorous tests
of SPA models than have previously been possible.

\acknowledgements This work benefited from valuable discussions with
our colleagues in a NASA Living-With-a-Star Focused-Science-Topic team
working on ``Flare Particle Acceleration Near the Sun and Contribution
to Large SEP Events.''  This work was supported in part by NASA grants
NNX07AP65G, NNX11AJ37G, and NNX12AB27G, DOE grant DE-FG02-07-ER46372,
and NSF grant AGS-1258998.

\appendix

\vspace{0.2cm} 
\section{Analytic Expression for the Wave Damping Rate Allowing for Relativistic Particles}
\label{appen:reldamping} 
\vspace{0.2cm}

We follow the standard approach in quasilinear theory of treating the plasma 
as infinite and homogeneous. To obtain Fourier transforms of the
fluctuating quantities, we define
the ``windowed'' Fourier transform
\begin{equation}
\tilde{g}(\bm{k}) = \frac{1}{(2\pi)^3} \int d^3 x\,\, 
g(\bm{x}) H(\bm{x}) \exp{(-i \bm{k}\cdot \bm{x})},
\label{eq:fourier1}
\end{equation}
where 
\begin{equation}
H(\bm{x})  = \left\{ \begin{array}{cc}\displaystyle 1 & \mbox{ if $|x|<\displaystyle \frac{L}{2}$, 
 $|y|<\displaystyle \frac{L}{2}$,  and
 $|z|<\displaystyle \frac{L}{2}$ 
} \vspace{0.2cm} \\
0 & \mbox{otherwise}
\end{array} \right. .
\label{eq:H}
\end{equation}
As mentioned in Section~\ref{sec:model}, our
Fourier-transform convention is the same as that of \cite{Stix92}, except
that we have an extra factor of $(2\pi)^{-3/2}$ on the right-hand side
of Equation~(\ref{eq:fourier1}). 
Accounting for this difference, we can use Eq.~(67) of Chapter~4 of
\cite{Stix92} to write the 
wave energy density~$\epsilon_w$ in
the form
\begin{equation}
\varepsilon_w = \lim_{L\rightarrow \infty}   \left(\frac{2\pi}{L}\right)^3 \int d^3k \,2W_k,
\label{eq:epsw} 
\end{equation} 
where $W_k$ is defined in Equation~(\ref{eq:defW}).

Wave-particle interactions cause the 
particle kinetic energy density of species~$s$ to change at the rate
\begin{equation}
\dot{K}_s =  \int d^3 p\,\, [(p^2 c^2+m^2_s c^4)^{1/2}-m_s c^2]
\left(\frac{\partial f_s}{\partial t}\right)_{\rm res},
\label{:heat_rate} 
\end{equation}
where $(\partial f_s/\partial t)_{\rm res}$ is given in
Equation~(\ref{eq:QLT0}).
The second term in brackets in Equation~(\ref{:heat_rate}), $m_s
c^2$, can be dropped, because $\int d^3p (\partial f_s/\partial
t)_{\rm res} = 0$.
Equation~(\ref{eq:epsw}) implies that wave-particle interactions cause
the wave energy density to change at the rate
\begin{equation}
\dot{\varepsilon}_w =  \left(\frac{2\pi}{L}\right)^3 
\int d^3 k \, 4\gamma_k W_{\bm{k}},
\end{equation}
where $\gamma_k$ is the imaginary part of the wave frequency.
Because the sum of the wave and particle-kinetic-energy densities is conserved,
\begin{equation}
\sum_s \dot{K}_s + \dot{\varepsilon}_w = 0.
\label{:e_conserved}
\end{equation}
Upon substituting Equation~(\ref{eq:QLT0}) into the
right-hand side of Equation~(\ref{:heat_rate}), integrating by parts,
and using the identity 
\begin{equation}
\frac{\partial v_\perp}{\partial p_\parallel} = \frac{\partial v_\parallel }{\partial p_\perp},
\label{eq:ID1} 
\end{equation} 
we can rewrite Equation~(\ref{:e_conserved})  in the form
\begin{equation}
\int d^3 k W_k I_k = 0,
\label{eq:econ2} 
\end{equation} 
where
\begin{equation}
I_k = \gamma_k - \sum_s \sum_{n=-\infty}^\infty \frac{\pi^2 q_s^2}{2}
\int_0^\infty dp_\perp \int_{-\infty}^\infty dp_\parallel
\frac{p_\perp^2 c^2}{\sqrt{p^2 c^2 + m_s^2 c^4}} \frac{|\psi_{n,k}^{(s)}|^2}{W_k}
 \delta(\omega_{kr} -
k_\parallel v_\parallel - n \Omega_s) G f_s.
\label{eq:defI} 
\end{equation} 
Equation~(\ref{eq:econ2}) must be satisfied for any function~$W_k$,
and hence $I_k$ must vanish at all~$k$. The condition that $I_k = 0$ at
each~$k$ reflects the fact that the change in the energy of the waves
within any small volume~$V$ of wavenumber space is equal and opposite
to the change in the particle kinetic energy that results from wave-particle
interactions involving waves with~$\bm{k} \in V$. We make use of this
fact when we evaluate~$\gamma_k$ in our numerical
calculations. Specifically, when we evaluate~$(\partial f_{\rm
  e}/\partial t)_{\rm res}$, we keep track of the change in particle
kinetic energy that results from interactions with waves within each
grid cell in wavenumber space. We use the term~$\Delta K_{i}$ to
denote the change in particle kinetic energy resulting from waves in
the $i^{\rm th}$ grid cell. We then evaluate~$\gamma_k$ within
the~$i^{\rm th}$ cell by setting $ (2\pi/L)^3 4\gamma_k W_k (\Delta
k)_i^3 \Delta t$ within that cell equal to~$-\Delta K_i$, where
$(\Delta k)_i^3$ is the volume of the grid cell in $k$~space, and
$\Delta t$ is the time step. By using this procedure, we ensure that
the changes in~$f_{\rm e}$ and~$F_k$ that result from wave-particle
interactions conserve energy to machine accuracy.

From the equation~$I_k=0$, we obtain
\begin{equation}
\gamma_k = \sum_s \sum_{n=-\infty}^\infty \frac{\pi^2 q_s^2}{2}
\int_0^\infty dp_\perp \int_{-\infty}^\infty dp_\parallel
\frac{p_\perp^2 c^2}{\sqrt{p^2 c^2 + m_s^2 c^4}} \frac{|\psi_{n,k}^{(s)}|^2}{W_k}
 \delta(\omega_{kr} -
k_\parallel v_\parallel - n \Omega_s) G f_s.
\label{eq:rel_damp}
\end{equation}
In the non-relativistic limit, Equation~(\ref{eq:rel_damp})  can be written
in the form
\begin{equation}
\gamma_{\bm{k}}=\sum_s \sum_{n=-\infty}^\infty \frac{\pi \omega^2_{ps}}{8 n_{\rm e}} \left |\frac{1}{k_{\|}}\right| \int^{\infty}_0 d v_\perp v^2_\perp \int^{\infty}_{-\infty} d v_\parallel \frac{|\psi_{n,\bm{k}}^{(s)} | ^2}{W_{\bm{k}}}\delta\left (v_\| -\frac{\omega_{kr}-n\Omega}{k_\|}\right) G_v f_s^{(v)},
\label{eq:rel_damp3}
\end{equation}
where $\omega_{ps} = \sqrt{4\pi n_s q_s^2/m_s}$ is the plasma
frequency of species~$s$, $f_s^{(v)} = f_s/m_{\rm e}^3$ is (in the
non-relativistic limit) the usual velocity-space distribution
function, and
\begin{equation}
G_v=\left(1-\frac{k_\parallel v_\parallel}{\omega_{kr}}\right)\frac{\partial}{\partial v_\perp}
+\left ( \frac{k_\parallel v_\perp}{\omega_{kr}}\right)\frac{\partial}{\partial v_\parallel}.
\label{:Gv}
\end{equation} 
Equation~(\ref{eq:rel_damp3}) is exactly the result derived by
\cite{kennel67}. Equation~(\ref{eq:rel_damp})  can thus be viewed as
a generalization of Kennel \& Wong's~(1967) result that allows for
relativistic particles.

\vspace{0.2cm} 
\section{Estimating the Error in our Approximate Collision Operator}
\label{sec:b} 
\vspace{0.2cm}

The Coulomb collision operator involves the quantities
\begin{equation}
K_1(\bm{p}) = \int d^3 p' f_{\rm e} (\bm{p}') u^{-1}, 
\end{equation} 
and
\begin{equation}
K_2(\bm{p}) = \int d^3 p' f_{\rm e} (\bm{p}') u, 
\end{equation} 
where $u = |\bm{p}'-\bm{p}|$.
To evaluate these integrals would require a very large number of operations per time step.  
In order to increase the speed of the calculations, we replace $K_1(\bm{p})$ and $K_2(\bm{p})$, respectively, with
\begin{equation}
H_1(\bm{p}) = \int d^3 p' f_{\rm M} (\bm{p}') u^{-1} 
\end{equation} 
and
\begin{equation}
H_2(\bm{p}) =\int d^3 p' f_{\rm M} (\bm{p}') u, 
\end{equation} 
where $f_{\rm M}$ is the Maxwellian distribution that has the same total particle
kinetic energy as $f_{\rm e}$.
In this case, $H_1(\bm{p})$ and $H_2(\bm{p})$ can be pre-calculated and depend
only on 
$p_\perp$, $p_\parallel$, $n_{\rm e}$, and~$T_{\rm e}$.

\begin{figure}[t]
\centerline{\includegraphics[width=12.cm]{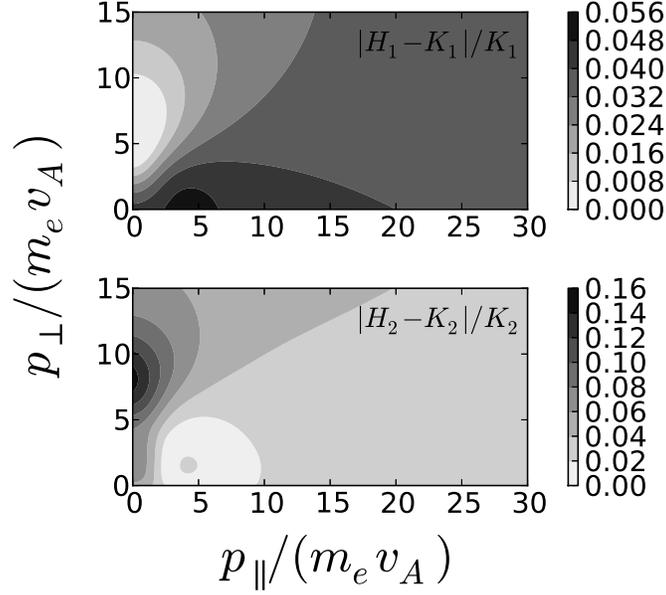}}
\caption{Difference between $H_1$ and $K_1$ (top panel) and $H_2$ and
  $K_2$ (bottom panel) in model solution~B at $t = 6.9\times 10^7
  \Omega^{-1}_p$.
  \label{fig:em}}\vspace{0.5cm}
\end{figure}

To estimate the error introduced by this approximation we compare
$H_1(\bm{p})$ to $K_1(\bm{p})$ and $H_2(\bm{p})$ to $K_2(\bm{p})$ in
numerical calculations A3, B, C, and D.  The maximum values of
$|H_1-K_1|/K_1$ and $|H_2-K_2|/K_2$ increase as $f_{\rm e}$ deviates
from a Maxwellian shape.  They increase and reach their maximum value
approximately at $t=t_{\rm inj}$. In model solutions A3 and D, which
have finite injection times, the maximum values of $|H_1-K_1|/K_1$ are
0.06 and 0.004, and the maximum values of $|H_2-K_2|/K_2$ are 0.18 and
0.02, respectively. In numerical calculations B and C, the maximum
value of $|H_1-K_1|/K_1$ is 0.06 and the maximum value of
$|H_2-K_2|/K_2$ is 0.16 within the time period we
consider. The plots of $|H_1-K_1|/K_1$ and $|H_2-K_2|/K_2$ from
numerical calculation~B at $t = 6.9\times 10^7 \Omega^{-1}_p$ are
shown in Figure \ref{fig:em}. The reason these errors are not much
larger is that most of the electrons in the numerical calculations
remain at low energies, where $f_{\rm e}$ is approximately Maxwellian.

\vspace{0.2cm} 
\section{Fast-Wave  Turbulence}
\label{appen:wave} 
\vspace{0.2cm} In many situations involving turbulence, fluctuations
or waves are excited at some large scale $\sim 1/k_0$ and then cascade
to smaller scales.  In our model, this large-scale excitation is
represented by the term $S_k$ in Equation~(\ref{eq:dFdt}), where
\begin{equation}
\dot{E}_0 = \frac{1}{2}\int d^3 k S_k, 
\end{equation}
is the total energy injection rate.  Since $S_k \propto
e^{-k^2/k^2_0}$ in our model, wave injection is limited to small
wavenumbers $\lesssim k_0$. If dissipation is only effective at
wavenumbers exceeding some dissipation wavenumber $k_d$, then
wavenumbers $\gg k_0$ and $\ll k_d$ are said to be in the inertial
range of the turbulence. In steady state, the energy cascade rate in
the inertial range must equal $\dot{E}_0$. From
Equation~(\ref{eq:turb}), when waves reach a steady state in
which $F_k = A_1 g_{\theta} k^{-7/2}$, the energy cascade rate per
solid angle per unit mass density is given by \citep{Chandran05},
\begin{equation}
\epsilon = \frac{9 \pi^2 c_2 A_1^2 g^2_\theta \sin^2\theta}{16 v_{\rm A}},
\label{eq:cas_angle_mass}
\end{equation}
where $c_2 \simeq 26.2$ is defined following
Equation~(\ref{eq:d_ttd}).  The quantity~$g_\theta$ is a function
of~$\theta$ that depends on the angular distribution of the input
power at $k\lesssim k_0$.  Equating the total energy injection rate
with the total cascade power, we obtain
\begin{equation}
\dot{E}_0 = 2\pi \int_0^\pi \epsilon \sin \theta d\theta.
\label{eq:valEdot} 
\end{equation} 
If waves
are injected isotropically (as in numerical calculation A4, B, C, and D,
in which $\epsilon$ is independent of
$\theta$), then $g_\theta = 1/\sin\theta$ 
\citep{Chandran05}, and
\begin{equation}
\dot{E} = \frac{9 \pi^3 c_2 A_1^2}{4 v_{\rm A}}
\label{eq:r1}
\hspace{1cm} \mbox{(isotropic $\epsilon$)}.
\end{equation}
On the other hand, in our numerical calculations A1, A2, and A3, we take 
$S_k \propto \sin\theta$, which leads to~$g_\theta = 1$ \citep{Chandran05}.
Equation~(\ref{eq:valEdot}) then yields 
\begin{equation}
\dot{E} = \frac{3 \pi^3 c_2 A^2_1}{2 v_{\rm A}}
\label{eq:r2}
\hspace{1cm} \mbox{(isotropic $F_k$)}.
\end{equation}
We can compare {\bf Equation~(\ref{eq:r2})} to the cascade rate implied by the equation 3.3 in MLM96,
\begin{equation}
\dot{E} = \frac{14 \pi^2 A^2_1}{v_{\rm A}}.
\label{eq:rmiller}
\hspace{1cm} \mbox{(MLM96)}
\end{equation}
The cascade rate in our model in Equation~(\ref{eq:r2}) is larger than
MLM96's by a factor of $3\pi c_2/28 = 8.8$ for a fixed isotropic
$F_k$. Therefore, if $\dot{E}$ is the same in our model and MLM96's,
then $F_k$ be will smaller in our model by a factor of~$\simeq 3$.

\bibliography{articles}

\begin{thebibliography}{43}
\expandafter\ifx\csname natexlab\endcsname\relax\def\natexlab#1{#1}\fi

\bibitem[{{Aschwanden}(2007)}]{Aschwanden07}
{Aschwanden}, M.~J. 2007, \apj, 661, 1242

\bibitem[{{Benz}(2008)}]{Benz08}
{Benz}, A.~O. 2008, Living Reviews in Solar Physics, 5, 1

\bibitem[{{Brown}(1971)}]{Brown71}
{Brown}, J.~C. 1971, \solphys, 18, 489

\bibitem[{{Carmichael}(1964)}]{Carmichael64}
{Carmichael}, H. 1964, NASA Special Publication, 50, 451

\bibitem[{{Caspi} \& {Lin}(2010)}]{Caspi10}
{Caspi}, A., \& {Lin}, R.~P. 2010, \apjl, 725, L161

\bibitem[{{Chandran}(2003)}]{chandran03}
{Chandran}, B.~D.~G. 2003, 599, 1426

\bibitem[{{Chandran}(2005)}]{Chandran05}
---. 2005, Physical Review Letters, 95, 265004

\bibitem[{{Chandran}(2008)}]{chandran08}
---. 2008, Physical Review Letters, 101, 235004

\bibitem[{Cho \& Lazarian(2002)}]{Cho02}
Cho, J., \& Lazarian, A. 2002, Physical Review Letters, 88, 245001

\bibitem[{{Eichler}(1979)}]{Eichler79}
{Eichler}, D. 1979, \apj, 229, 413

\bibitem[{Ginzburg(1960)}]{ginz60}
Ginzburg, V.~L. 1960, The Propagation of Electromagnetic Waves in Plasmas
  (Oxford: Pergammon)

\bibitem[{{Grigis} \& {Benz}(2004)}]{Grigis04}
{Grigis}, P.~C., \& {Benz}, A.~O. 2004, \aap, 426, 1093

\bibitem[{{Hirayama}(1974)}]{Hirayama74}
{Hirayama}, T. 1974, \solphys, 34, 323

\bibitem[{{Ishikawa} {et~al.}(2011){Ishikawa}, {Krucker}, {Takahashi}, \&
  {Lin}}]{Ishikawa11}
{Ishikawa}, S., {Krucker}, S., {Takahashi}, T., \& {Lin}, R.~P. 2011, \apj,
  737, 48

\bibitem[{Kennel \& Engelmann(1966)}]{Kennel66}
Kennel, C., \& Engelmann, F. 1966, Phys. Fluids, 9, 2377

\bibitem[{{Kennel} \& {Wong}(1967)}]{kennel67}
{Kennel}, C.~F., \& {Wong}, H.~V. 1967, Journal of Plasma Physics, 1, 75

\bibitem[{{Kopp} \& {Holzer}(1976)}]{Kopp76}
{Kopp}, R.~A., \& {Holzer}, T.~E. 1976, \solphys, 49, 43

\bibitem[{{Krucker} {et~al.}(2010){Krucker}, {Hudson}, {Glesener}, {White},
  {Masuda}, {Wuelser}, \& {Lin}}]{Krucker10}
{Krucker}, S., {Hudson}, H.~S., {Glesener}, L., {White}, S.~M., {Masuda}, S.,
  {Wuelser}, J.-P., \& {Lin}, R.~P. 2010, \apj, 714, 1108

\bibitem[{{Krucker} {et~al.}(2011){Krucker}, {Hudson}, {Jeffrey}, {Battaglia},
  {Kontar}, {Benz}, {Csillaghy}, \& {Lin}}]{Krucker11}
{Krucker}, S., {Hudson}, H.~S., {Jeffrey}, N.~L.~S., {Battaglia}, M., {Kontar},
  E.~P., {Benz}, A.~O., {Csillaghy}, A., \& {Lin}, R.~P. 2011, \apj, 739, 96

\bibitem[{{Lin} {et~al.}(1981){Lin}, {Schwartz}, {Pelling}, \&
  {Hurley}}]{Lin81}
{Lin}, R.~P., {Schwartz}, R.~A., {Pelling}, R.~M., \& {Hurley}, K.~C. 1981,
  \apjl, 251, L109

\bibitem[{{Lin} {et~al.}(2003){Lin}, {Krucker}, {Hurford}, {Smith}, {Hudson},
  {Holman}, {Schwartz}, {Dennis}, {Share}, {Murphy}, {Emslie}, {Johns-Krull},
  \& {Vilmer}}]{Lin03}
{Lin}, R.~P., {et~al.} 2003, \apjl, 595, L69

\bibitem[{{Liu} {et~al.}(2009){Liu}, {Petrosian}, \& {Mariska}}]{Liu09}
{Liu}, W., {Petrosian}, V., \& {Mariska}, J.~T. 2009, \apj, 702, 1553

\bibitem[{{Lynn} {et~al.}(2012){Lynn}, {Parrish}, {Quataert}, \&
  {Chandran}}]{lynn12}
{Lynn}, J.~W., {Parrish}, I.~J., {Quataert}, E., \& {Chandran}, B.~D.~G. 2012,
  ArXiv e-prints

\bibitem[{{Lynn} {et~al.}(2013){Lynn}, {Quataert}, {Chandran}, \&
  {Parrish}}]{lynn13}
{Lynn}, J.~W., {Quataert}, E., {Chandran}, B.~D.~G., \& {Parrish}, I.~J. 2013,
  \apj, 777, 128

\bibitem[{{Lynn} {et~al.}(2014){Lynn}, {Quataert}, {Chandran}, \&
  {Parrish}}]{lynn14}
---. 2014, ArXiv e-prints

\bibitem[{{Miller} {et~al.}(1996){Miller}, {Larosa}, \& {Moore}}]{Miller96}
{Miller}, J.~A., {Larosa}, T.~N., \& {Moore}, R.~L. 1996, \apj, 461, 445

\bibitem[{{Miller} {et~al.}(1997){Miller}, {Cargill}, {Emslie}, {Holman},
  {Dennis}, {LaRosa}, {Winglee}, {Benka}, \& {Tsuneta}}]{Miller97}
{Miller}, J.~A., {et~al.} 1997, \jgr, 102, 14631

\bibitem[{{Petrosian} {et~al.}(2006){Petrosian}, {Yan}, \&
  {Lazarian}}]{Petrosian06}
{Petrosian}, V., {Yan}, H., \& {Lazarian}, A. 2006, \apj, 644, 603

\bibitem[{{Pongkitiwanichakul} {et~al.}(2012){Pongkitiwanichakul}, {Chandran},
  {Karpen}, \& {DeVore}}]{Peera12a}
{Pongkitiwanichakul}, P., {Chandran}, B.~D.~G., {Karpen}, J.~T., \& {DeVore},
  C.~R. 2012, \apj, 757, 72

\bibitem[{{Priest} \& {Forbes}(2000)}]{Priest2000}
{Priest}, E., \& {Forbes}, T. 2000, Irish Astronomical Journal, 27, 235

\bibitem[{{Rosenbluth} {et~al.}(1957){Rosenbluth}, {MacDonald}, \&
  {Judd}}]{rosenbluth57}
{Rosenbluth}, M.~N., {MacDonald}, W.~M., \& {Judd}, D.~L. 1957, Physical
  Review, 107, 1

\bibitem[{{Schlickeiser} \& {Miller}(1998)}]{Schlickeiser98}
{Schlickeiser}, R., \& {Miller}, J.~A. 1998, \apj, 492, 352

\bibitem[{{Selkowitz} \& {Blackman}(2004)}]{Selkowitz04}
{Selkowitz}, R., \& {Blackman}, E.~G. 2004, \mnras, 354, 870

\bibitem[{{Shalchi} {et~al.}(2004){Shalchi}, {Bieber}, {Matthaeus}, \&
  {Qin}}]{shalchi04a}
{Shalchi}, A., {Bieber}, J.~W., {Matthaeus}, W.~H., \& {Qin}, G. 2004, \apj,
  616, 617

\bibitem[{{Shalchi} \& {Schlickeiser}(2004)}]{shalchi04b}
{Shalchi}, A., \& {Schlickeiser}, R. 2004, \aap, 420, 799

\bibitem[{{Stix}(1992)}]{Stix92}
{Stix}, T.~H. 1992, {Waves in plasmas}

\bibitem[{{Tsuneta}(1996)}]{Tsuneta96}
{Tsuneta}, S. 1996, \apj, 456, 840

\bibitem[{{Tsuneta} {et~al.}(1992){Tsuneta}, {Hara}, {Shimizu}, {Acton},
  {Strong}, {Hudson}, \& {Ogawara}}]{Tsuneta92}
{Tsuneta}, S., {Hara}, H., {Shimizu}, T., {Acton}, L.~W., {Strong}, K.~T.,
  {Hudson}, H.~S., \& {Ogawara}, Y. 1992, \pasj, 44, L63

\bibitem[{van~de Vorst(2003)}]{Vorst03}
van~de Vorst, H. 2003, Iterative Krylov Methods for Large Linear Systems
  (Cambridge: Cambridge Univ. Press)

\bibitem[{{Yan} \& {Lazarian}(2004)}]{Yan04}
{Yan}, H., \& {Lazarian}, A. 2004, \apj, 614, 757

\bibitem[{{Yan} \& {Lazarian}(2008)}]{yan08b}
---. 2008, \apj, 673, 942

\bibitem[{{Yan} {et~al.}(2008){Yan}, {Lazarian}, \& {Petrosian}}]{Yan08}
{Yan}, H., {Lazarian}, A., \& {Petrosian}, V. 2008, \apj, 684, 1461

\bibitem[{Zakharov \& Sagdeev(1970)}]{Zakharov70}
Zakharov, V.~E., \& Sagdeev, R.~Z. 1970, Sov. Phys. Dokl., 15, 439

\end{thebibliography}
\end{document}